\documentclass[11pt,showpacs,preprint,preprintnumbers,nofootinbib,superscriptaddress,amsmath,amssymb]{revtex4}

\usepackage{graphicx}
\usepackage{dcolumn}
\usepackage{bm}
\usepackage{amssymb}
\usepackage{amsmath}
\usepackage{epsfig}    
\usepackage{color}
\usepackage{slashed}

\begin{document} 

\title{Enhancement of the $H^\pm W^\mp Z$ vertex in the three scalar doublet model}

%
\author{Stefano Moretti}
\email{S.Moretti@soton.ac.uk}
\affiliation{School of Physics and Astronomy, University of Southampton, Southampton, SO17 1BJ, United Kingdom}
\author{Diana Rojas}
\email{D.Rojas@soton.ac.uk}
\affiliation{School of Physics and Astronomy, University of Southampton, Southampton, SO17 1BJ, United Kingdom}
\affiliation{Instituto de F\'isica, Benem\'erita Universidad Aut\'onoma de Puebla, \\
Apdo. Postal J-48, C.P. 72570 Puebla, Puebla, M\'exico}
\author{Kei Yagyu}
\email{K.Yagyu@soton.ac.uk}
\affiliation{School of Physics and Astronomy, University of Southampton, Southampton, SO17 1BJ, United Kingdom}

\begin{abstract}

We compute one-loop induced trilinear vertices with physical charged Higgs bosons $H^\pm$ and ordinary gauge bosons, i.e., 
$H^\pm W^\mp Z$ and $H^\pm W^\mp \gamma$, in the model with  
two active plus one inert scalar doublet fields under a $Z_2(\text{unbroken})\times \tilde{Z}_2(\text{softly-broken})$ symmetry. 
The $Z_2$ and $\tilde{Z}_2$ symmetries are introduced to guarantee the stability of a dark matter candidate and to forbid the flavour 
changing neutral current at the tree level, respectively. 
The dominant form factor $F_Z$ of the $H^\pm W^\mp Z$ vertex can be enhanced by non-decoupling effects of extra scalar boson loop contributions. 
We find that, in such a model, $|F_Z|^2$ can be one order of magnitude larger than that predicted in two Higgs doublet models
under the constraints from vacuum stability, perturbative unitarity and the electroweak precision observables. 
In addition, the branching fraction of the $H^\pm \to W^\pm Z$ $(H^\pm \to W^\pm \gamma)$ mode
can be of order 10~(1)\% level when the mass of $H^\pm$ is below the top quark mass. 
Such a light $H^\pm$ is allowed by the so-called Type-I and Type-X Yukawa interactions which appear under the classification of 
the $\tilde{Z}$ charge assignment of the quarks and leptons. 
We also calculate the cross sections for the processes $H^\pm \to W^\pm Z$ and $H^\pm \to W^\pm \gamma$ onset by  the top quark decay $t\to H^\pm b$
and electroweak $H^\pm$ production at the LHC. 

\end{abstract}
\maketitle

\section{Introduction}

Although the discovery of the Standard Model (SM) like Higgs boson at the 
Large Hadron Collider (LHC)~\cite{Higgs_coupling1,Higgs_coupling2,Higgs_coupling3,Higgs_coupling4}  
suggests that there is an isospin doublet scalar field in the Higgs sector, the possibility of the existence more Higgs doublets still remains open. 
In fact, 
a second doublet is often introduced in new physics models such as the Minimal Supersymmetric SM (MSSM)~\cite{HHG}. 
In addition, models with a multi-doublet structure have also been discussed based upon various physics motivations, e.g., to explain 
tiny neutrino masses via radiative generation~\cite{Zee}, to provide a dark matter (DM) candidate~\cite{IDM} 
and to supply extra CP violating phases~\cite{CPV} for the explanation of the baryon asymmetry of Universe. 
Thus, testing the existence of additional doublet fields is quite important to probe new physics scenarios beyond the SM.

One of the most important features of models with  multi-Higgs doublets  is the appearance of physical extra scalar bosons such as charged Higgs bosons $H^\pm$. 
In particular, the properties of $H^\pm$ states strongly depend on the structure of the Higgs sector, e.g.,
the symmetries of the model, the actual number of doublets, the mass spectrum, etc. 
Therefore, through the detection of $H^\pm$ and by measuring those properties, e.g., the mass, couplings, production cross sections and decay rates,  
one can directly probe the existence of additional doublets as well attempt extracting the structure of the Higgs sector. 

Among the various observables related to $H^\pm$, studying the $H^\pm W^\mp Z$  vertex is quite interesting because of the following features.  
Firstly, it has been known that the $H^\pm W^\mp Z$ vertex does not appear at the tree level\footnote{The $H^\pm W^\mp \gamma$ vertex does also not 
appear at the tree level in any models with the $U(1)_{\text{em}}$ symmetry.} in  multi-doublet models~\cite{Grifols}, because 
of an approximate global $SU(2)$ symmetry known as the custodial symmetry\footnote{In fact, the custodial symmetry is broken by the $U(1)_Y$ coupling in the kinetic sector which generates
the mass difference between the $W$ and $Z$ bosons. } 
in the kinetic terms for the doublet fields. 
Secondly, although the $H^\pm W^\mp Z$ vertex is loop induced, its magnitude can be enhanced by non-decoupling effects of particles running in the loop, especially  for the case where they come from the sector which breaks the custodial symmetry.  
For example, 
the top and bottom quark loop contributions to the $H^\pm W^\mp Z$ vertex give the quadratic dependence upon the top quark mass~\cite{f}, which 
is responsible for the violation of the custodial symmetry in the Yukawa sector. 
In Refs.~\cite{b1,b2}, the impact of extra Higgs boson loop contributions on the $H^\pm W^\mp Z$ vertex
has been evaluated in the 2-Higgs Doublet Model (2HDM) \cite{HHG}.  
It has been shown that a large mass splitting between the CP-odd Higgs boson and the charged one gives a sizable correction to the $H^\pm W^\mp Z$ vertex. 
From the above reasons, it is clear that the strength of the $H^\pm W^\mp Z$ vertex measures the effects of the violation of the custodial symmetry in the model
embedding it. Therefore, by measuring this vertex, we can indirectly observe such a new physics effect. 

Feasibility studies to measure the $H^\pm W^\mp Z$ vertex have  been performed in Ref.~\cite{HWZ_LHC} for the LHC and in Ref.~\cite{Yanase} for future linear colliders. 

In this paper, we calculate the magnitude of the $H^\pm W^\mp V$ ($V=Z, \gamma$) vertices at the one-loop level 
in the 3-Higgs Doublet Model (3HDM), in which the Higgs sector is composed of 
two $active$ (with a non-zero Vacuum Expectation Value (VEV)) and one $inert$ (without a non-zero VEV) doublet scalar fields.
In this model, the scalar bosons from the inert doublet field give an additional contribution to the $H^\pm W^\mp V$ vertex with respect to the 
top/bottom quarks and scalar bosons from the active doublet loop contributions. 
As a phenomenological application, 
we also discuss how such new contributions change the decay branching fractions of the $H^\pm \to W^\pm Z$ and $H^\pm \to W^\pm \gamma$ modes and, 
consequently, the production cross sections involving these decay processes at the LHC. 

This paper is organized as follows. 
In Sec.~\ref{sec:model}, we define the Lagrangian of the 3HDM, i.e., the scalar potential and the Yukawa interactions. 
In Sec.~\ref{sec:HWZ}, we introduce the form factors of the $H^\pm W^\mp V$ vertices and discuss relationships 
between these form factors and effective operators. We then explain how to calculate these form factors at the one-loop level. 
In Sec.~\ref{sec:const}, we summarise various constraints on the parameters of our model. 
From the theoretical point of view, we consider vacuum stability and perturbative unitarity. 
As experimental constraints, we take into account the bounds from the
Electro-Weak (EW) $S$, $T$ and $U$ parameters, the flavour experiments and direct searches for $H^\pm $ states from LEP-II and the LHC Run-I. 
In Sec.~\ref{sec:results}, we show numerical results for the form factors of the $H^\pm W^\pm V$ vertices, branching fractions of $H^\pm$ and 
their signal cross sections at the LHC.  Our conclusion is given in Sec.~\ref{sec:conclusion}.
In Appendix, we present the full analytic expressions for the form factors of the $H^\pm W^\mp V$ vertices. 

\section{The Model}\label{sec:model}

\begin{table}[t]
\begin{center}
{\renewcommand\arraystretch{1.2}
\begin{tabular}{c|cccccccc||ccc}\hline\hline
&
\multicolumn{8}{c||}{$(Z_2,~\tilde{Z}_2)$ charge}
&\multicolumn{3}{c}{Mixing factor}\\     \cline{2-12}
&$\Phi_1$&$\Phi_2$&$\eta$&$Q_L$&$L_L$&
$u_R$&$d_R$&$e_R$
&$\xi_u$ &$\xi_d$&$\xi_e$ \\\hline
Type-I &$(+,+)$&
$(+,-)$&$(-,+)$&$(+,+)$&$(+,+)$&
$(+,-)$&$(+,-)$&$(+,-)$&$\cot\beta$&$\cot\beta$&$\cot\beta$ \\\hline
Type-II&$(+,+)$&
$(+,-)$&$(-,+)$&$(+,+)$&$(+,+)$&
$(+,-)$
&$(+,+)$&$(+,+)$& $\cot\beta$&$-\tan\beta$&$-\tan\beta$ \\\hline
Type-X &$(+,+)$&
$(+,-)$&$(-,+)$&$(+,+)$&$(+,+)$&
$(+,-)$
&$(+,-)$&$(+,+)$&$\cot\beta$&$\cot\beta$&$-\tan\beta$ \\\hline
Type-Y &$(+,+)$&
$(+,-)$&$(-,+)$&$(+,+)$&$(+,+)$&
$(+,-)$
&$(+,+)$&$(+,-)$& $\cot\beta$&$-\tan\beta$&$\cot\beta$ \\\hline\hline
\end{tabular}}
\caption{Charge assignments of the unbroken $Z_2$ symmetry and the 
softly-broken $\tilde{Z}_2$ symmetry.  
The mixing factors in the Yukawa interaction terms in Eq.~(\ref{yukawa_thdm}) are also shown.}
\label{yukawa_tab}
\end{center}
\end{table}

We give a brief review of the 3HDM\footnote{The model with two inert plus one active doublets 
have been discussed in Refs.~\cite{2plus1_1,2plus1_2}.} of which 
the Higgs sector is composed of two active and one inert isospin doublet scalar fields~\cite{idm2,3hdm_vs}. 
We represent the active doublets as $\Phi_1$ and $\Phi_2$ whereas the inert doublet as $\eta$. 
Such an inert nature can be realised by assuming an unbroken $Z_2$ symmetry in the scalar potential, 
in which only $\eta$ has an odd parity while all the other fields are assigned to be even.    
One of the important consequences of imposing such a $Z_2$ symmetry is  
that the lightest neutral scalar component in $\eta$ can be a DM candidate, because it cannot decay into SM particles. 

In addition to the $Z_2$ symmetry, we impose another $Z_2$ symmetry, denoted by $\tilde{Z}_2$ to distinguish it from the above one, 
which is required to forbid the Flavour Changing Neutral Current (FCNCs) at the tree level. 
This prescription is the same as that in the 2HDM~\cite{GW}. 
For the $\tilde{Z}_2$ symmetry, we consider the softly-broken case, since  
avoidance of tree level FCNCs can already be achieved in this case. 
Under the $\tilde{Z}_2$ symmetry, four independent types of Yukawa interactions (Type-I, -II, -X and -Y)~\cite{4types,4types2,typeX}
are allowed depending on the assignment of the $\tilde{Z}_2$ charge to the SM fermions. 
In Tab.~\ref{yukawa_tab}, we show the charge assignments required by the $Z_2$ and $\tilde{Z}_2$ symmetries
for the three scalar doublets $\Phi_1$, $\Phi_2$ and $\eta$ and all the SM fermions, where 
$L_L~(e_R^{})$ is the left (right)-handed lepton doublet (singlet) and 
$Q_L~(u_R^{},~d_R^{})$ is the left (right)-handed quark doublet (up-type and down-type quark singlets).

\subsection{The scalar potential}

The most general scalar potential under the $SU(2)_L\times U(1)_Y\times Z_2\times \tilde{Z}_2$ symmetry is given by
\begin{align}
V(\Phi_1,\Phi_2,\eta) &=\mu_\eta^2 \eta^\dagger \eta +  \mu_1^2 \Phi_1^\dagger \Phi_1 + \mu_2^2 \Phi_2^\dagger \Phi_2 - (\mu_3^2 \Phi_1^\dagger \Phi_2 + \text{h.c.}) \notag\\
 & +\frac{1}{2}\lambda_1 (\Phi_1^\dagger \Phi_1)^2 +\frac{1}{2}\lambda_2 (\Phi_2^\dagger \Phi_2)^2  +\lambda_3(\Phi_1^\dagger \Phi_1)(\Phi_2^\dagger \Phi_2)
+\lambda_4|\Phi_1^\dagger \Phi_2|^2 +\frac{1}{2}[\lambda_5(\Phi_1^\dagger \Phi_2)^2 + \text{h.c.} ] \notag\\
& +\frac{1}{2}\lambda_\eta (\eta^\dagger \eta)^2  + \rho_1(\Phi_1^\dagger\Phi_1)(\eta^\dagger\eta)
 +\rho_2|\Phi_1^\dagger \eta|^2
+\frac{1}{2}[\rho_3(\Phi_1^\dagger \eta)^2 + \text{h.c.} ]\notag\\
&+ \sigma_1(\Phi_2^\dagger\Phi_2)(\eta^\dagger\eta)
 +\sigma_2|\Phi_2^\dagger \eta|^2
+\frac{1}{2}[\sigma_3(\Phi_2^\dagger \eta)^2 + \text{h.c.} ], \label{pot}
\end{align}
where $\mu_3^2$, $\lambda_5$, $\rho_3$ and $\sigma_3$ are complex parameters in general. 
Throughout the paper, we take these parameters to be real for simplicity.  
The scalar fields can be parameterised as 
\begin{align}
\Phi_i = \left[
\begin{array}{cc}
w_i^+ \\ 
\frac{1}{\sqrt{2}}(h_i+v_i+iz_i)
\end{array}\right],~~(i=1,2), \quad  
\eta = \left[
\begin{array}{cc}
\eta^+ \\ 
\frac{1}{\sqrt{2}}(\eta_H^{}+i\eta_A^{}) 
\end{array}\right],
\end{align}
where $v_i$ are the VEVs of $\Phi_i$ with $v_1^2+v_2^2=v^2\simeq (246$ GeV$)^2$. 
The ratio of the two VEVs is parameterized as the usual way by $\tan\beta =v_2/v_1$. 

The mass formulae for the active sector are exactly the same as those in the 2HDM at the tree level. 
The mass eigenstates for the active scalar bosons are given as: 
\begin{align}
&\begin{pmatrix}
w_1^\pm \\
w_2^\pm 
\end{pmatrix}
= R(\beta)
\begin{pmatrix}
G^\pm \\
H^\pm 
\end{pmatrix},~
\begin{pmatrix}
z_1^{} \\
z_2^{}
\end{pmatrix}
= R(\beta)
\begin{pmatrix}
G^0 \\
A 
\end{pmatrix},
\begin{pmatrix}
h_1^{} \\
h_2^{}
\end{pmatrix}
= R(\alpha)
\begin{pmatrix}
H \\
h 
\end{pmatrix},\notag\\
&R(\theta) = 
\begin{pmatrix}
\cos\theta & -\sin\theta \\
\sin\theta & \cos\theta
\end{pmatrix}, 
\end{align}
where $G^\pm$ and $G^0$ are the Nambu-Goldstone (NG) bosons which are absorbed as their longitudinal components by 
the $W^\pm$ and $Z$ bosons, respectively.  We define the $h$ state to be the SM-like Higgs boson with a mass of about 125 GeV discovered at the LHC. 

The squared masses of the $H^\pm$ and $A$ states are then calculated as 
\begin{align}
m_{H^\pm}^2=M^2-\frac{v^2}{2}(\lambda_4+\lambda_5),\quad m_A^2&=M^2-v^2\lambda_5, \label{massch}
\end{align}
where 
\begin{align}
M^2 = \frac{\mu_3^2}{\sin\beta\cos\beta}. 
\end{align}
The squared masses for the CP-even scalar states and the mixing angle $\alpha$ are expressed by
\begin{align}
&m_H^2=\cos^2(\alpha-\beta)M_{11}^2+\sin^2(\alpha-\beta)M_{22}^2+\sin2(\alpha-\beta)M_{12}^2, \label{mass2} \\
&m_h^2=\sin^2(\alpha-\beta)M_{11}^2+\cos^2(\alpha-\beta)M_{22}^2-\sin2(\alpha-\beta)M_{12}^2,\\
&\tan 2(\alpha-\beta)=\frac{2M_{12}^2}{M_{11}^2-M_{22}^2},  \label{tan2a}
\end{align}
where $M_{ij}^2$ ($i,j=1,2$) are the mass matrix elements in the basis of $(h_1',h_2')$ defined in Eq.~(\ref{Higgs_base}):
\begin{align}
M_{11}^2&=v^2(\lambda_1\cos^4\beta+\lambda_2\sin^4\beta)+\frac{v^2}{2}(\lambda_3+\lambda_4+\lambda_5)\sin^22\beta,\notag\\
M_{22}^2&=M^2+v^2\sin^2\beta\cos^2\beta\left[\lambda_1+\lambda_2-2(\lambda_3+\lambda_4+\lambda_5)\right],\notag\\
M_{12}^2&=\frac{v^2}{2}\sin2\beta(-\lambda_1\cos^2\beta+\lambda_2\sin^2\beta)+\frac{v^2}{2}\sin2\beta\cos2\beta(\lambda_3+\lambda_4+\lambda_5). \label{mateven}
\end{align}

Because of the unbroken $Z_2$ symmetry, the scalar bosons from $\eta$ do not mix with those from $\Phi_1$ and $\Phi_2$. 
Thus, the mass formulae of the inert scalar bosons are simply given by 
\begin{align}
m_{\eta^\pm}^2 & = \mu_\eta^2 + \frac{v^2}{2}\left[\rho_1  \cos^2\beta + \sigma_1 \sin^2\beta \right], \\ 
m_{\eta_H^{}}^2 & = \mu_\eta^2 + \frac{v^2}{2}\left[(\rho_1 + \rho_2 + \rho_3)\cos^2\beta+(\sigma_1 + \sigma_2 + \sigma_3)\sin^2\beta \right], \\
m_{\eta_A^{}}^2 & = \mu_\eta^2 + \frac{v^2}{2}\left[(\rho_1 + \rho_2 - \rho_3)\cos^2\beta+(\sigma_1 + \sigma_2 - \sigma_3)\sin^2\beta \right]. 
\end{align}

\subsection{The Yukawa Lagrangian}

The most general form under the $\tilde{Z}_2$ symmetry is given by  
\begin{align}
-{\mathcal L}_Y =
&Y_{u}{\overline Q}_Li\sigma_2\Phi^*_uu_R^{}
+Y_{d}{\overline Q}_L\Phi_dd_R^{}
+Y_{e}{\overline L}_L\Phi_e e_R^{}+\text{h.c.},
\end{align}
where $\Phi_{u,d,e}$ are $\Phi_1$ or $\Phi_2$. 
The interaction terms  are expressed in terms of mass eigenstates
of the Higgs bosons as
\begin{align}
-{\mathcal L}_Y^{\text{int}}=&
\sum_{f=u,d,e}\frac{m_f}{v}\left( \xi_h^f{\overline
f}fh+\xi_H^f{\overline f}fH-2iI_f \xi_f{\overline f}\gamma_5fA\right)\notag\\
&+\frac{\sqrt{2}}{v}\left[V_{ud}\overline{u}
\left(m_d\xi_d\,P_R-m_u\xi_uP_L\right)d\,H^+
+m_e\xi_e\overline{\nu^{}}P_Re^{}H^+
+\text{h.c.}\right],  \label{yukawa_thdm}
\end{align}
where $I_f$ is the third component of the isospin for a fermion $f$. 
In Eq.~(\ref{yukawa_thdm}),  $\xi_h^f$ and $\xi_H^f$ are defined by
\begin{align}
\xi_h^f &= \sin(\beta-\alpha)+\xi_f \cos(\beta-\alpha), \label{hff} \\
\xi_H^f &= \cos(\beta-\alpha)-\xi_f \sin(\beta-\alpha), 
\end{align}
and $\xi_f$ in each type of Yukawa interactions are listed in Tab.~\ref{yukawa_tab}. 

It is important to mention here that there is the so-called SM-like limit or alignment limit defined by $\sin(\beta-\alpha)\to 1$~\cite{Gunion-Haber,Dev}. 
In this limit, all the $h$ coupling constants to the SM particles 
become the same values as those of the SM values. 
In fact, the ratios of  $hf\bar{f}$ and $hVV$ couplings in our model to those in the SM are respectively
given as $\xi_h^f$ given in Eq.~(\ref{hff}) and $\sin(\beta-\alpha)$.

\section{The $H^\pm W^\mp V$ vertex}\label{sec:HWZ}

The amplitude of $H^\pm \to W^\pm V$ ($V= Z,~\gamma$) is expressed as 
\begin{align}
i\mathcal{M}(H^\pm\to W^\pm V) = igm_W V_{V}^{\mu\nu}\epsilon_{W\mu}(p_W)\epsilon_{V\nu}(p_V), ~~ \text{for} ~~ V= Z,~\gamma, 
\end{align}
where $V_{V}^{\mu\nu}$ is written in terms of the following three dimensionless form factors: 
\begin{align}
V_{V}^{\mu\nu} = g^{\mu\nu}F_V +\frac{p_V^\mu p_W^\nu}{m_W^2}G_V +
i\epsilon^{\mu\nu\rho\sigma} \frac{p_{V\rho}^{} p_{W\sigma}^{}}{m_W^2} H_V,  \label{HWZ1}
\end{align}
with $p_W^\mu$ and $p_V^\mu$ being the incoming momenta for $W^\pm$ and $V$, respectively. 
For the case of $V=\gamma$, the Ward identity guarantees the following relation; 
\begin{align}
V_{\gamma}^{\mu\nu}p_{\gamma\nu} = 0. 
\end{align}
From this relation, the form factor $F_\gamma$ is written as 
\begin{align}
F_\gamma = \frac{G_\gamma}{2} \left(1-\frac{m_{H^\pm}^2}{m_W^2}\right), 
\end{align}
where we use $p_W^2=m_W^2$ and $(p_W^{}+p_\gamma)^2=m_{H^\pm}^2$. 

In our model, the $H^\pm W^\mp V$ vertices do not appear at the tree level, just like in the 2HDM. 
This is clearly seen by introducing the so-called Higgs basis of the active scalar doublets defined as
\begin{align}
\left(\begin{array}{c}
\Phi_1 \\ 
\Phi_2
\end{array}\right)
= R(\beta) 
\left(\begin{array}{c}
\Phi \\ 
\Psi
\end{array}\right), \label{Higgs_basis}
\end{align}
where 
\begin{align}
\Phi = \left[
\begin{array}{cc}
G^+ \\ 
\frac{1}{\sqrt{2}}(h_1'+v+iG^0)
\end{array}\right], \quad  
\Psi = \left[
\begin{array}{cc}
H^+ \\ 
\frac{1}{\sqrt{2}}(h_2'+iA)
\end{array}\right],  \label{Higgs_base}
\end{align}
with $h_1'=H\cos(\beta-\alpha)+h\sin(\beta-\alpha)$ and 
$h_2'=-H\sin(\beta-\alpha)+h\cos(\beta-\alpha)$. 
The kinetic Lagrangian for $\Phi_1$ and $\Phi_2$ is then rewritten as 
\begin{align}
{\cal L}_{\text{kin}} &= |D_\mu \Phi_1|^2 + |D_\mu \Phi_2|^2 
=  |D_\mu \Phi|^2 + |D_\mu \Psi|^2,  
\end{align}
where $D_\mu$ is the covariant derivative. 
Since the gauge-gauge-scalar type vertex is proportional to the Higgs VEV $v$, these 
vertices come from the $|D_\mu \Phi|^2$ term as only $\Phi$ has a non-zero VEV. 
However, the physical charged Higgs bosons $H^\pm$ are contained in the $|D_\mu \Psi|^2$ term. 
Therefore, the $H^\pm W^\mp Z$ vertex is absent at the tree level\footnote{If we consider models 
which contain scalar fields with isospin larger than 1/2 such as triplets, 
the $H^\pm W^\mp Z$ vertex can appear at tree level. 
The expression for the $H^\pm W^\mp Z$ vertex can be found in Refs.~\cite{Grifols,Yanase} in the general extended Higgs sector
which contains Higgs multiplets with the isospin $T$ and the hypercharge $Y$. 
In addition, it has been known that 
in models with an extension of the gauge sector such as $SU(2)\times SU(2)\times U(1)$~\cite{abe}, 
the $H^\pm W^\mp Z$ vertex also appears at the tree level.  }. 
The above statement can be generalised to a model with $N$ active doublet scalar fields. 
In that case, we can also define a base transformation similar to the one of Eq.~(\ref{Higgs_basis}). 
Regarding the $H^\pm W^\mp \gamma$ vertex, it does not appear at  tree level 
in any models based on the $SU(2)_L\times U(1)_Y\to U(1)_{\text{em}}$ gauge theory, 
because of the $U(1)_{\text{em}}$ invariance and the consequent Ward identity. 

The form factors defined in Eq.~(\ref{HWZ1}) are introduced from the following effective Lagrangian~\cite{f,b1}:
\begin{align}
\mathcal{L}_{\text{eff}} = f_{Z}^{} H^+W^-_\mu Z^\mu +
g_{V}^{} H^+F_W^{\mu\nu}F_{V\mu\nu}+ih_{V}^{}\epsilon_{\mu\nu\rho\sigma} H^+F_W^{\mu\nu}F_V^{\rho\sigma}+ \text{h.c.},  \label{effective}
\end{align}
where $F_W^{\mu\nu}$ and $F_V^{\mu\nu}$ are the field strength tensors for $W^\pm$ and $V$, respectively. 
It can be seen that the coefficient $f_{Z}$ has  mass dimension one
whereas $g_{H^\pm WV}$ and $h_Z$ have mass dimension minus one. 
Hence, the coefficient $f_{Z}$ can be proportional to a squared mass $(M_i^2)$ of a particle running in the loop according to a  dimensional analysis:
\begin{align}
f_{Z} \sim  gg_Z^{}\frac{M_i^2}{v}{\cal F}(M_i^2), 
\end{align}
where ${\cal F}$ is a dimensionless function. 
Typically, it is expressed by the logarithmic function of $M_i^2$. 
On the other hand,  
$g_{Z}$ and $h_{Z}$ can be expressed as 
\begin{align}
g_{Z},~ h_{Z} \sim \frac{gg_Z^{}}{v}{\cal G}(M_i^2), 
\end{align}
where ${\cal G}$ is another dimensionless function of $M_i^2$. 
Therefore, only the coefficient $f_{H^\pm WZ}$ can be enhanced significantly due to the $M_i^2$ dependence, 
so that the form factor $F_{Z}$ gives the dominant contribution to the $H^\pm W^\mp Z$ vertex. 
In fact, it has been pointed out in Ref.~\cite{f} that the top/bottom loop contribution to the form factor $F_{Z}$
is proportional to $m_t^2$ only, as $m_t\gg m_b$. 
The origin of the quadratic dependence can be understood in terms of  
the Yukawa coupling $H^+ t\bar{b}$, which is proportional to $m_t/v$ as in Eq.~(\ref{yukawa_thdm}), and of
another $m_t$ coming from the chirality flipped effect. 
Similarly, the quadratic mass dependence appears in the extra Higgs boson loop contribution as discussed in Ref.~\cite{b1}. 
This too can be understood,  as the trilinear $H^\pm S S'$ ($S$ and $S'$ being extra scalar bosons) couplings can be  
rewritten by squared masses of extra scalar bosons. 

Another important reason for the appearance of a $M_i^2$ dependence in $F_{Z}$ is 
in relation to a violation of the custodial $SU(2)_V$ symmetry. 
As it has been discussed in Ref.~\cite{b1}, 
the dimension three term in Eq.~(\ref{effective}) comes from the following operator\footnote{The operator $\text{Tr}[D_\mu {\bm \Phi} D^\mu {\bm \Psi}]$ also gives 
the $H^\pm W^\mp Z$ term in the effective Lagrangian which is proportional to $\sin^2\theta_W$. 
However, such an effect is cancelled by the counter term of the $H^\pm WZ$ vertex. } 
\begin{align}
\text{Tr}[ \sigma_3 (D_\mu {\bm \Phi})^\dagger D^\mu {\bm \Psi}], \label{custodial}
\end{align}
where ${\bm \Phi}=(\Phi^c,\Phi)$ and ${\bm \Psi}=(\Psi^c,\Psi)$ with 
$\Phi^c=i\sigma_2 \Phi^*$ and $\Psi^c=i\sigma_2 \Psi^*$ 
are the $2\times 2$ representation form of the Higgs doublets. 
They are translated under the $SU(2)_L\times SU(2)_R$ symmetry by ${\bm \Phi}\to U_L{\bm \Phi}U_R^\dagger$ and 
${\bm \Psi}\to U_L{\bm \Psi}U_R^\dagger$, where $U_L$ and $U_R$ are respectively the $SU(2)_L$ and $SU(2)_R$ 
unitary transformation matrices. 
We can see that the operator given in Eq.~(\ref{custodial}) is not invariant under the $SU(2)_R$ transformation, so that 
this operator breaks the $SU(2)_R$ invariance. 
Since the custodial $SU(2)_V$ symmetry corresponds to the remaining symmetry after the EW symmetry breaking, i.e., 
$SU(2)_L\times SU(2)_R\to SU(2)_V$ and a violation of the $SU(2)_R$ symmetry means a violation of the $SU(2)_V$ symmetry. 

Therefore, the quadratic mass dependence in $F_Z$ can be understood as a result of the custodial symmetry breaking. 
In fact, 
it has been known that the mass difference between the top and bottom quarks gives the violation of the custodial symmetry 
in the Yukawa sector. 
In addition,  that between $A$ and $H^\pm$ 
also gives the violation of the custodial symmetry in the Higgs potential~\cite{Pomarol}. 
Since the top quark mass is already known by experiments, the top quark loop contribution to the $H^\pm W^\mp Z$ vertex is determined by its mass\footnote{In our model, 
the top quark loop contribution also depends on $\tan\beta$, 
and in all the four types of Yukawa interactions, its dependence is given by $\cot\beta$. }. 
In contrast, parameters in the scalar sector have not yet determined by experiments except for the Higgs boson mass of about 125 GeV, 
so that we can expect a sizable 
enhancement of the $H^\pm W^\mp Z$ vertex from scalar boson loop effects in suitable regions of the 3HDM parameter space. 

In the following, we discuss how we calculate the form factors of the $H^\pm W^\mp V$ vertices. 
We can separately consider the one-loop contributions to the vertices from the 1PI diagrams and the counter terms as 
\begin{align}
(F_V,G_V,H_V) = (F_V^{\text{1PI}}+\delta F_V,~G_V^{\text{1PI}}+\delta G_V,~H_V^{\text{1PI}}+\delta H_V), 
\end{align}
where $X_V^{\text{1PI}}$ and $\delta X_V$ are respectively the 1PI and the counter term contributions to the form factor $X_V$ ($X=F,~G$ and $H$).  
Their analytic expressions are given in App.~A.

The counter term contributions are obtained as follows.  
First, we define the renormalized two point function for the $W^\pm$-$H^\pm$ mixing as 
\begin{align}
\hat{\Gamma}_{WH}^\mu(p^2) = (-ip^\mu)\hat{\Gamma}_{WH}(p^2), 
\end{align}
where $p^\mu$ is the incoming four momentum of $H^\pm$. 
The renormalised form factor $\hat{\Gamma}_{WH}$ is given by 
\begin{align}
\hat{\Gamma}_{WH}(p^2) = im_W \delta_{GH} + \Gamma_{WH}^{\text{1PI}}(p^2),  
\end{align}
where $\delta_{GH}$ is the counter term for the $G^\pm$-$H^\pm$ mixing,  
and $\Gamma_{WH}^{\text{1PI}}$ is the 1PI diagram contribution to the $W^\pm$-$H^\pm$ mixing.  
The analytic expression of $\Gamma_{WH}^{\text{1PI}}$ is given in App.~A. 
The counter term is obtained by the shift of the charged NG boson field $G^\pm$:
\begin{align} 
G^\pm \to (1+\delta Z_{G}/2)G^\pm + \delta_{GH}H^\pm. 
\end{align}
By imposing the on-shell renormalisation condition~\cite{KOSY,finger2}
\begin{align}
\hat{\Gamma}_{WH}(p^2=m_{H^\pm}^2) = 0, \label{ward}
\end{align}
we can determine the counter term 
\begin{align}
\delta_{GH} &=
 i\frac{\Gamma_{WH}^{\text{1PI}}(p^2=m_{H^\pm}^2)}{m_W}. \label{del}
\end{align}
We then obtain the counter term contribution to the $H^\pm W^\mp V$ vertex as 
\begin{align}
\mathcal{L}_{GWV}& = -\frac{g}{c_W^{}}m_W^{}s_W^2 G^+W^-_\mu Z^\mu+em_WG^+W^-A_\mu+\text{h.c.} \notag\\ 
& \to  -\frac{g}{c_W^{}} m_W s_W^2 \delta_{GW}H^+W_\mu^- Z^\mu +em_W  \delta_{GW}H^+W_\mu^- A^\mu  +\cdots  \label{delGW}, 
\end{align}
where $s_W^{}=\sin\theta_W$ and $c_W^{}=\cos\theta_W$ with $\theta_W$ being the weak mixing angle. 
From Eqs.~(\ref{del}) and (\ref{delGW}), $\delta F_V$ is given by 
\begin{align}
\delta F_Z = -i\frac{s_W^2}{c_W^{}}\frac{\Gamma_{WH}^{\text{1PI}}(p^2=m_{H^\pm}^2)}{m_W}, \quad 
\delta F_\gamma = is_W\frac{\Gamma_{WH}^{\text{1PI}}(p^2=m_{H^\pm}^2)}{m_W}.   \label{delFF}
\end{align}

We then obtain the finite results for the form factors of  the $H^\pm W^\mp Z$ and $H^\pm W^\mp\gamma$ vertices. 
In the case of $\sin(\beta-\alpha)=1$, $m_{H}^{}=m_{A}\gg m_{H^\pm}$ and $m_{\eta_H^{}}^{}=m_{\eta_A^{}}\gg m_{\eta^\pm}$, we obtain 
\begin{align}
F_Z \simeq \frac{\cot\beta}{16\pi^2 v^2 c_W^{}}\left[N_c m_t^2 
+\frac{M^2-m_A^2}{2}(\tan^2\beta -1)+\left(m_{\eta_A^{}}^2-m_{\eta^\pm}^2-\frac{v^2}{2}\rho_2\right)\right], \label{fz_app}
\end{align}
where the first, second and third terms correspond to the contributions from $t$-$b$, active and inert scalar boson loops, respectively. 
From the above expression, we can clearly see the quadratic mass dependences $m_t^2$, $m_{A}^2$ and $m_{\eta_A^{}}^2$. 
However, as it will be discussed in the next section, 
the case considered in the above, i.e.,  $m_{H}^{}=m_{A}\gg m_{H^\pm}$ and $m_{\eta_H^{}}^{}=m_{\eta_A^{}}\gg m_{\eta^\pm}$ also gives 
the similar quadratic dependence in the EW $T$ parameter. 
Therefore, too large mass difference between $H^\pm$ and $A$ (with $m_H^{}=m_A^{}$) and that between 
$\eta^\pm$ and $\eta_A^{}$ (with $m_{\eta_H^{}}=m_{\eta_A^{}}$) are not allowed. 
Instead of taking the above case, 
we can consider the case with $\sin(\beta-\alpha)=1$, $m_{A}\gg m_{H^\pm}(=m_H^{})$ and  $m_{\eta_A^{}}\gg m_{\eta^\pm}(=m_{\eta_H^{}})$, where 
the contribution to the $T$ parameter from extra scalar boson loops is cancelled. 
We then obtain 
\begin{align}
F_Z \simeq \frac{\cot\beta}{16\pi^2 v^2 c_W^{}}\left[N_c m_t^2 +(M^2-m_{H^\pm}^2)(\tan^2\beta-1) F\left(\frac{m_{H^\pm}^2}{m_A^2}\right)-\frac{v^2}{2}(\rho_2+\rho_3)
F\left(\frac{m_{\eta^\pm}^2}{m_{\eta_A}^2}\right) 
\right], \label{fz_app}
\end{align}
where $N_c=3$ is the color factor, and the function $F$ is given by 
\begin{align}
F(r) = -\frac{1}{4(1-r)^2}\left[3-4r+r^2+2(2-r)r\ln r\right]-\frac{1}{2}\ln r. 
\end{align}
This function has the following asymptotic behavior:
\begin{align}
F(r)\simeq -\frac{3}{4}-\frac{1}{2}\ln r ~~\text{for}~~r\ll1,\quad 
F(r)\simeq -\frac{1}{4}~~\text{for}~~r\gg1,\quad
F(r)\simeq \frac{1-r}{2}~~\text{for}~~r \simeq 1. 
\end{align}
In this case, although the quadratic dependence $m_A^2$ and $m_{\eta_A}^2$ disappears, there still remains their logarithmic dependence.   

\section{Constraints}\label{sec:const}

\subsection{Vacuum stability}

The stability condition for the Higgs potential is given by requiring that 
the Higgs potential is bounded from below in any direction of the scalar boson space. 
The necessary and sufficient condition to guarantee such a positivity of the potential has been derived in Ref.~\cite{3hdm_vs} as
\begin{align}
& \lambda_\eta > 0,\quad \lambda_1 > 0,\quad \lambda_2 > 0, \\
& \sqrt{\lambda_1\lambda_2} + \bar{\lambda} >0, \quad
  \sqrt{\lambda_\eta\lambda_1} + \bar{\rho} >0, \quad 
  \sqrt{\lambda_\eta\lambda_2} + \bar{\sigma} >0, \label{vs2}  \\
& \sqrt{\lambda_\eta}\bar{\lambda} + \sqrt{\lambda_1}\bar{\sigma} + \sqrt{\lambda_2}\bar{\rho}>0 \text{~~or~~}
\lambda_\eta\bar{\lambda}^2 + \lambda_1\bar{\sigma}^2 + \lambda_2\bar{\rho}^2 -\lambda_\eta\lambda_1\lambda_2-2\bar{\lambda}\bar{\rho}\bar{\sigma}<0, 
 \label{vs3}  
\end{align}
\vspace*{-1.2cm}
\begin{align}
& \bar{\lambda} = \lambda_3 + \text{MIN}(0,~\lambda_4+\lambda_5,~\lambda_4-\lambda_5), \notag\\
&\bar{\rho} = \rho_1+ \text{MIN}(0,~\rho_2+\rho_3,~\rho_2-\rho_3), \notag\\
&\bar{\sigma} = \sigma_1+ \text{MIN}(0,~\sigma_2+\sigma_3,~\sigma_2-\sigma_3). 
\end{align}

\subsection{Unitarity}

Some combinations of scalar quartic couplings are constrained from perturbative unitarity. 
In the 3HDM, the $s$ wave amplitude matrix for all the 2-to-2 body scalar boson elastic scatterings have been 
calculated in Ref.~\cite{Moretti-Yagyu} in the high energy limit. 
We obtain the following  independent eigenvalues or sub-matrices for the $s$ wave amplitude matrix as
\begin{align}
X_1 &= 
\begin{pmatrix} 
3\lambda_\eta &2\rho_1 + \rho_2 & 2\sigma_1 + \sigma_2  \\
2\rho_1 + \rho_2 & 3\lambda_1 & 2\lambda_3 + \lambda_4\\
 2\sigma_1 + \sigma_2 & 2\lambda_3 + \lambda_4 & 3\lambda_2 
\end{pmatrix}, ~
X_2  =
\begin{pmatrix} 
\lambda_\eta &\rho_2 & \sigma_2  \\
\rho_2 & \lambda_1 &  \lambda_4\\
\sigma_2 & \lambda_4 & \lambda_2 
\end{pmatrix},~
X_3 = 
\begin{pmatrix} 
\lambda_\eta &\rho_3 & \sigma_3  \\
\rho_3 & \lambda_1 &  \lambda_5\\
\sigma_3 & \lambda_5 & \lambda_2 
\end{pmatrix}, \label{3times3}  
\end{align}
\begin{align}
y_1^\pm & = \lambda_3 + 2\lambda_4 \pm 3\lambda_5, \label{y1}\\
y_2^\pm & = \rho_1 + 2\rho_2 \pm 3\rho_3, \\
y_3^\pm & = \sigma_1 + 2\sigma_2 \pm 3\sigma_3, \\
y_4^\pm & = \lambda_3 \pm \lambda_5, \\
y_5^\pm & = \rho_1 \pm \rho_3, \\
y_6^\pm & = \sigma_1 \pm \sigma_3, \\
y_7^\pm & =  \lambda_3 \pm \lambda_4 , \\
y_8^\pm &=  \rho_1 \pm \rho_2 , \\
y_9^\pm & =  \sigma_1 \pm \sigma_2.  \label{y9}
\end{align}
We then require the following condition:
\begin{align}
|x_i|     & < 8\pi,\quad |y_j^\pm|  < 8\pi,~~(i,j=1,...9) ,  
\end{align}
where $x_i$ are the eigenvalues of $X_1$, $X_2$ and $X_3$. 

\subsection{$S$, $T$ and $U$ parameters}

The EW oblique parameters $S$, $T$ and $U$~\cite{stu} 
can be modified from the SM prediction 
by the extra scalar boson loop contributions and the modified SM-like Higgs boson couplings. 
The differences in the predictions of the $S$, $T$ and $U$ parameters in the 3HDM and those in the SM
are given in the case with $\sin(\beta-\alpha)= 1$, $m_H^{}=m_A^{}$, $m_{\eta_H^{}}^{}=m_{\eta^\pm}^{}$ as 
\begin{align}
\Delta T& \simeq \frac{1}{24\pi^2 \alpha_{\text{em}} v^2}(m_{H^\pm}^{}-m_A^{})^2, \label{delt}\\
\Delta U& \simeq \frac{1}{12\pi}\Big(\ln\frac{m_A^2}{m_{H^\pm}^2}+\frac{2m_{H^\pm}^{}}{m_A}-2 \Big)\simeq 0, 
\end{align}
assuming $m_A^{}\simeq m_{H^\pm}^{}$, and
\begin{align} 
\Delta S& \simeq \frac{1}{12\pi}\Big(\ln \frac{m_{A}^2}{m_{H^\pm}^2}+ \ln \frac{m_{\eta_A}^2}{m_{\eta^\pm}^2} -\frac{5}{6}\Big),~~\text{for}~~m_{\eta_A^{}}\gg m_{\eta^\pm}, \label{dels_1}\\
\Delta S& \simeq \frac{1}{12\pi}\Big(\ln \frac{m_{A}^2}{m_{H^\pm}^2} -\frac{5}{6}\Big),~~\text{for}~~m_{\eta^\pm}\gg m_{\eta_A^{}}, \label{dels_2} \\
\Delta S& \simeq \frac{1}{12\pi}\Big(\ln \frac{m_{A}^2}{m_{H^\pm}^2} +\frac{m_{\eta_A^{}}^{}}{m_{\eta^\pm}^{}}-1\Big)\simeq 0,~~\text{for}~~m_{\eta^\pm}\simeq m_{\eta_A^{}}.  \label{dels_deg}
\end{align}
The general expression is given in Ref.~\cite{Moretti-Yagyu}. 
From the global fit of the EW precision data, 
$\Delta S$ and $\Delta T$ are extracted by fixing  $\Delta U=0$  as 
\begin{align}
\Delta S =0.05\pm 0.09,\quad  \Delta T = 0.08\pm 0.07,
\label{STallowed}
\end{align}
with the correlation coefficient of +0.91~\cite{Baak:2012kk}. 

\begin{figure}[t]
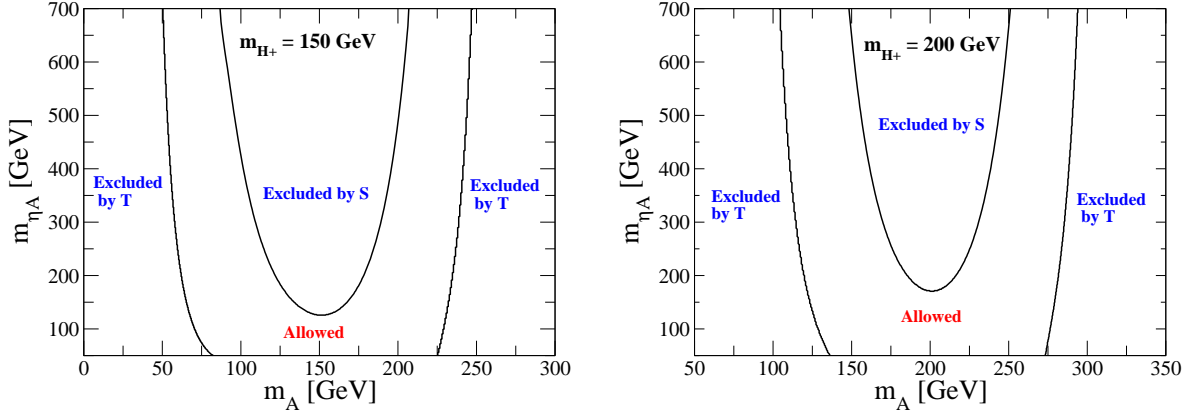

\begin{center}
\includegraphics[width=75mm]{stu_150.eps}\hspace{5mm}
\includegraphics[width=75mm]{stu_200.eps} 
\caption{Constraint from the $S$ and $T$ parameters on the $m_A^{}$-$m_{\eta_A}$ plane
in the case of $m_H=m_A$, $\sin(\beta-\alpha)=1$ and $m_{\eta^\pm}=m_{\eta_H}=m_A/2$. 
The charged Higgs boson mass is fixed to be $150$ GeV (left panel) and 200 GeV (right panel). 
The 95\% CL excluded regions are indicated in the figure.  }
\label{stu_fig}
\end{center}
\end{figure}

In Fig.~\ref{stu_fig}, we show the constraint from the $S$ and $T$ parameters on the $m_A^{}$-$m_{\eta_A}$ plane. 
We take $\sin(\beta-\alpha)=1$, $m_H^{}=m_A$ and $m_{\eta^\pm}^{}=m_{\eta_H^{}}=m_A^{}/2$, which is also taken in the numerical results shown in Sec.~\ref{sec:results}.  
In the left and right panel, $m_{H^\pm}$ is fixed to be 150 GeV and 200 GeV, respectively. 
We can see that, for $m_{\eta_A^{}}\simeq m_{\eta^\pm}$, a magnitude of the mass splitting between $A$ and $H^\pm$ 
to be larger than about 75 GeV is excluded by the $T$ parameter due to the quadratic dependence of the mass splitting shown in Eq.~(\ref{delt}). 
In this case, the contribution to $\Delta S$ is almost zero as it is seen in Eq.~(\ref{dels_deg}). 
Conversely, in the case of $m_{\eta_A^{}}^{}\gg m_{\eta^\pm}$, 
the positive logarithmic contribution to $\Delta S$ appears as shown in Eq.~(\ref{dels_1}) and  
a too large mass splitting between $\eta_A^{}$ and $\eta^\pm$ is excluded by $\Delta S$. 
However, the constraint from $\Delta S$ is getting milder when there is a positive contribution to $\Delta T$, because 
of the positive correlation between $\Delta S$ and $\Delta T$. 
Therefore, in order to have a large mass splitting between $\eta_A^{}$ and $\eta^\pm$, which is 
required to obtain a significant contribution to the $H^\pm W^\mp Z$ vertex, 
we need a mass splitting between $A$ and $H^\pm$.

\subsection{Flavour constraints}

We can apply the same constraints from the $B$ physics measurements as those in the 2HDM to our 3HDM, because of the same structure of the active sector. 
From the $b\to s\gamma$ process, 
the mass bound of $m_{H^\pm} \gtrsim 322$ GeV is given at 95\% confidence level (CL) in models with the Type-II and Type-Y Yukawa interactions
with $\tan\beta\gtrsim 2$ via the next-to-next-to-leading order calculation performed in Refs.~\cite{bsg_322,Misiak}. 
This bound is getting stronger when a smaller value of $\tan\beta$ is considered. 
In models with Type-I and Type-X Yukawa interactions, the constraint from $b\to s\gamma$ is only important in the small $\tan\beta$ case. 
For instance, the lower limit on $m_{H^\pm}$ is given to be 
about 100, 200 and 800 GeV at 95\% CL in the cases of $\tan\beta=2.5$, 2 and 1, respectively~\cite{Misiak}.  
 
The $B^0$-$\bar{B}^0$ mixing also gives a bound on $m_{H^\pm}$, especially for  small $\tan\beta$'s. 
In the case of $\tan\beta=1$, $m_{H^\pm}\lesssim 500$ GeV is excluded at 95\% CL in models with all the types of Yukawa interactions~\cite{Stal}, which 
is stronger than the constraint from $b\to s\gamma$ for the Type-II and Type-Y cases. 
This bound becomes rapidly weaker when we consider $\tan\beta\gtrsim 1$, e.g., 
for $\tan\beta=1.5~(2)$, the limit is $m_{H^\pm}\lesssim 300 ~(100)$ GeV at 95\% CL. 

\subsection{Direct search at LEP II}

At the LEP II experiment, charged Higgs bosons have been searched 
via the $e^+e^-\to Z^*/\gamma^* \to H^+ H^-$ process~\cite{LEPII}. 
From the non-observation of a significant excess, the lower mass limit 
has been taken to be about 80 GeV at 95\% CL under  
the assumption of BR$(H^\pm \to \tau^\pm \nu)$ + BR$(H^\pm \to cs)=1$. 
The slightly stronger bound $m_{H^\pm}\gtrsim 90$ GeV can be obtained assuming BR$(H^\pm \to \tau^\pm \nu)=1$. 

\subsection{Direct search at LHC Run-I}

\begin{figure}[t]
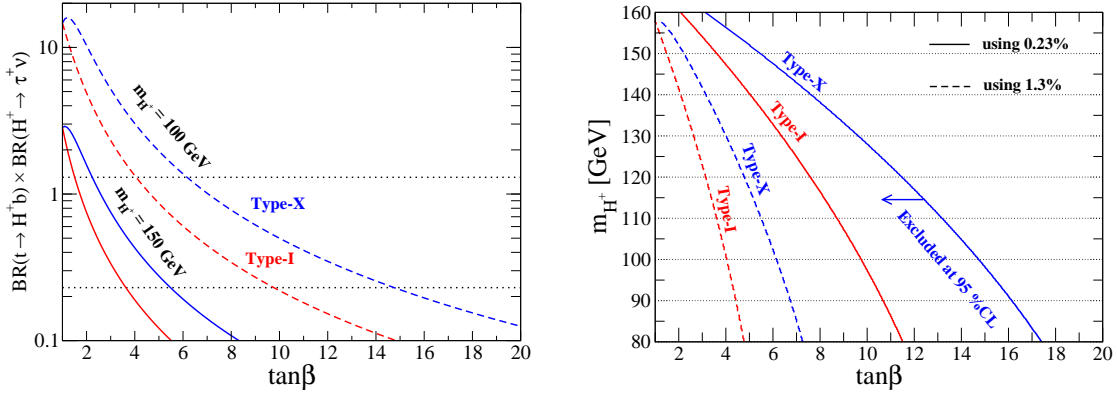

\begin{center}
\includegraphics[width=70mm]{top_decay.eps} \hspace{5mm}
\includegraphics[width=70mm]{contour_topdecay.eps}
\caption{(Left panel) 
The product of branching fractions BR($t\to H^+ b)\times$BR$(H^+\to \tau^+\nu)$ as a function of $\tan\beta$ in the Type-I (red curves) and Type-X (blue curves) 2HDMs/3HDMs. 
We take $m_{H^\pm}=m_A^{}=m_H=M$ and $\sin(\beta-\alpha)=1$ in this plot. 
The dashed and solid curves respectively show the cases of $m_{H^\pm}=$100 GeV and 150 GeV.
The horizontal dotted lines show the upper limits (0.23\% and 1.3\%) from the LHC data. 
(Right) Excluded parameter regions on the $\tan\beta$-$m_{H^\pm}$ plane in the Type-I and Type-X 2HDMs/3HDMs. 
Regions inside from each curve are excluded at 95\% CL by the measurement of top decay $t\to H^\pm b \to \tau^\pm b \nu$. 
The solid and dashed curves are the results using the upper limit on BR($t\to H^+ b)\times$BR$(H^+\to \tau^+\nu)$
to be 0.23\% and 1.3\%, respectively. 
}
\label{top_decay}
\end{center}
\end{figure}

At the LHC, $H^\pm$ searches have been performed for the two cases: the low mass region $m_{H^\pm}< m_t + m_b$ and the high mass region
$m_{H^\pm} >  m_t + m_b$. 
For the low mass case, the $t\to H^+ b$ decay is used as the $H^\pm$ production mode 
and the full process $pp\to t\bar{t}\to b\bar{b}H^\pm W^\mp$ with the $H^\pm \to \tau^\pm \nu$ decay
has thus been analysed. 
Using the data obtained at $\sqrt{s}=8$ TeV after 19.5 fb$^{-1}$ of the integrated luminosity, the upper limit on the product of branching ratios
BR$(t\to H^\pm b)\times$BR$(H^\pm \to \tau^\pm\nu)$ has been obtained to be between 0.23\% and 1.3\% at 95\% CL for $m_{H^\pm}$ in the range of 80 GeV to 160 GeV~\cite{ATLAS_H+}. 

In the left panel of Fig.~\ref{top_decay}, 
the above product of branching ratios is shown as a function of $\tan\beta$ in the Type-I and Type-X 2HDMs. 
Because the light $H^\pm$ scenario, i.e., $m_{H^\pm}<m_t$, in the Type-II and Type-Y 2HDMs has already been excluded by 
$b\to s\gamma$ data as explained in Sec.~IV-D, we here only show the Type-I and Type-X cases. 
In the Type-X 2HDM, the product of the branching fractions is slightly larger than that in the Type-I 2HDM. 
This can be understood in such a way that in the Type-X 2HDM the branching fraction of $H^\pm \to \tau^\pm \nu$ is 
enhanced as $\tan\beta$  is increased, while it does not depend on $\tan\beta$ in the Type-I 2HDM.
For example, BR$(H^+\to \tau^+\nu)$ can be almost 100\% when $\tan\beta \gtrsim 3$ in the Type-I 2HDM, but it is 
about 40\% in the Type-I 2HDM.   
In contrast, the branching ratio of $t\to H^+ b$ is given by the same value in both Type-I and Type-X 2HDMs. 
Therefore, a bit stronger bound on $\tan\beta$ for a fixed value of $m_{H^\pm}$ is obtained in the Type-X 2HDM. 
For example, if we use the stronger bound for BR$(t\to H^\pm b)\times$BR$(H^\pm \to \tau^\pm\nu)$, i.e., 0.23\%, 
$\tan\beta \lesssim 6~(4)$ and 15~(10) are excluded for $m_{H^\pm}=100$ and 150 GeV in the Type-X (Type-I) 2HDM.  

For the high $H^\pm$ mass region, i.e., $m_{H^\pm}>m_t$, 
the production process $gb \to tH^\pm$ (i.e., $H^\pm$-strahlung) can be used instead of the top quark 
decay\footnote{Notice that we have emulated both the top quark production and the decay as well as $H^\pm$-strahlung through the single $gg\to tbH^\pm$ mode, in the spirit of \cite{Guchait:2001pi}.}. 
The 95\% CL upper limit on the cross section times branching ratio $\sigma(pp\to tH^\pm+X)\times \text{BR}(H^\pm \to \tau^\pm \nu)$
has been given to be between 0.76 pb and 4.5 fb  in the range of $m_{H^\pm}=$180 GeV to 1 TeV~\cite{ATLAS_H+}.
This limit gives an upper limit on $\tan\beta$ for a fixed value of $m_{H^\pm}$ in the 2HDMs. 
For example, $\tan\beta \gtrsim 50~(60)$ at $m_{H^\pm}=200$ (230) GeV can be excluded at 95\% CL in the MSSM~\cite{ATLAS_H+}, where 
a similar bound is expected to be obtained in the Type-II 2HDM because of the same structure of the Yukawa interaction\footnote{In the Type-Y 2HDM, 
although the same production cross section of $pp \to tH^\pm+X$ is obtained as in the Type-II case, 
the branching fraction of $H^\pm \to \tau^\pm \nu$ is significantly suppressed due to the enhancement of the 
decay rate of the $H^\pm \to bc$ mode~\cite{cb}. 
Therefore, the bound in the Type-Y 2HDM can be much weaker than that in the Type-II case.  }. 
In the Type-I and Type-X 2HDMs, the production cross section of $pp \to tH^\pm+X$ 
is significantly suppressed by a factor $\cot^2\beta$, so that we cannot expect to obtain an important bound in the high mass region.

\subsection{Summary of the constraints on $m_{H^\pm}$}

In Tab.~\ref{const_tab}, we present the summary of the current experimental bounds on $m_{H^\pm}$ in the 2HDMs/3HDMs with the four types of Yukawa interactions from various 
experimental observations.  

\begin{table}[t]
\begin{center}
{\renewcommand\arraystretch{1.2}
\begin{tabular}{c|cccccccc||ccc}\hline\hline
Experiment&95\% CL lower lim. on $m_{H^\pm}$ &tan$\beta$ &Type&Comments \\  \hline
$b\to s\gamma $&322 GeV&-&II and Y& \\  \cline{2-5}
               &(800, 200, 100) GeV&(1, 2, 2.5)&I and X& \\  \hline
$B^0\text{-}\bar{B}^0 $&(500, 300, 100) GeV&(1, 1.5, 2)&All& \\  \hline
LEP II&(80, 90) GeV&-&All&${\cal B}_{\tau\nu}+{\cal B}_{cs}=1$,~${\cal B}_{\tau\nu}=1$  \\  \hline
$t \to H^\pm b $ &(160, 140, 100) GeV&(1,~2,~4)&I&Using 1.3\% (See Fig.~\ref{top_decay}) \\ \cline{2-5}
at the LHC Run-I&(160, 150, 130) GeV&(1,~2,~4)&X&Using 1.3\% (See Fig.~\ref{top_decay})  \\  \hline \hline
\end{tabular}}
\caption{The 95\% CL lower bound on $m_{H^\pm}$ in the 2HDMs/3HDMs from various experimental measurements for a fixed value of $\tan\beta$. 
For the row of LEP II, 80 (90) GeV is given for the case of ${\cal B}_{\tau\nu}+{\cal B}_{cs}=1$,~$({\cal B}_{\tau\nu}=1)$, where 
${\cal B}_{\tau\nu}$ and ${\cal B}_{cs}$ are the branching fractions of $H^\pm \to \tau^\pm\nu$ and $H^\pm \to cs$ modes, respectively. 
}
\label{const_tab}
\end{center}
\end{table}

\section{Numerical results}\label{sec:results}

In this section, we perform numerical evaluations for the $H^\pm W^\mp V$ vertices and related observables.  
In particular, we focus on the light $H^\pm$ case,  i.e, $m_{H^\pm}={\cal O}(100)$ GeV, because of its  phenomenological interest. 
As we discussed in Sec.~IV, such a scenario is allowed in the Type-I and Type-X Yukawa interactions from flavour constraints, 
so that we consider these types only in this section. 
First, we evaluate the form factors of the $H^\pm W^\mp Z$ and $H^\pm W^\mp \gamma$ vertices.
For the $H^\pm W^\mp \gamma$ vertex, since the form factor $F_\gamma$ is  related to $G_\gamma$ 
by the Ward identity, 
we only show $G_\gamma$ and $H_\gamma$. 
Second, we show all the branching fractions of $H^\pm$, 
including the $H^\pm \to W^\pm Z$ and $H^\pm \to W^\pm \gamma$ modes. 
Finally, we discuss cross sections for various signal processes involving the $H^\pm \to W^\pm Z$ and $H^\pm \to W^\pm \gamma$ decays at the LHC.

In our model, there are 16 independent parameters in the potential given in Eq.~(\ref{pot}), namely, 
$\mu_{1\text{-}3}^2$, $\mu_\eta^2$, $\lambda_{1\text{-}5}$, $\lambda_\eta$, $\rho_{1\text{-}3}$ and $\sigma_{1\text{-}3}$. 
They are divided into 8 parameters in the active sector ($\mu_{1\text{-}3}^2$ and $\lambda_{1\text{-}5}$) 
and the remaining 8 parameters ($\mu_\eta^2$, $\lambda_\eta$, $\rho_{1\text{-}3}$ and $\sigma_{1\text{-}3}$). 

After the tadpole conditions are imposed, 
the former 8 parameters can be expressed by $v$, $\tan\beta$, $\sin(\beta-\alpha)$ $m_h$, $m_H^{}$, $m_A^{}$, $m_{H^\pm}$ and $M^2$. 
Two of the 8 parameters, $v$ and $m_h$, should be used to reproduce the gauge boson masses and the observed Higgs boson mass, i.e., 
$v\simeq 246$ GeV and $m_h\simeq 125$ GeV. 
Furthermore, the Higgs boson search data at the LHC suggests that the observed Higgs boson is SM-like~\cite{Higgs_coupling1,Higgs_coupling2,Higgs_coupling3,Higgs_coupling4}, so that taking 
$\sin(\beta-\alpha)\approx1$ gives a good benchmark scenario as we explained in Sec.~II. 
We thus take $\sin(\beta-\alpha)=1$ in the following calculation. 

Regarding the latter 8 parameters, we proceed as follows.
First, we take $\lambda_\eta=0$, as this gives a four-point interaction among the inert scalar bosons that 
does not affect the following analysis.  
Second, we take $\rho_1$ and $\sigma_1$ so as to satisfy the vacuum stability condition given in Eqs.~(\ref{vs2}) and (\ref{vs3}) for given values of 
$\rho_{2,3}$ and $\sigma_{2,3}$:
\begin{align}
&\rho_1 = \text{MIN}(0,\rho_2+\rho_3,\rho_2-\rho_3),\quad \sigma_1 = \text{MIN}(0,\sigma_2+\sigma_3,\sigma_2-\sigma_3).  \label{bp1}
\end{align}
Finally, the remaining 5 parameters can be expressed in terms of three masses of the inert scalar bosons ($m_{\eta^\pm}$, $m_{\eta_A^{}}$ and $m_{\eta_H^{}}$)
and the $\rho_2$ and $\rho_3$ parameters. 
In this parametrisation, the $\sigma_2$ and $\sigma_3$ parameters are given as the outputs:
\begin{align}
\sigma_2 &= -\rho_2\cot^2\beta +\frac{1}{v^2\sin^2\beta}\left(m_{\eta_A^{}}^2+m_{\eta_H^{}}^2-2m_{\eta^\pm}^2\right), \\
\sigma_3 &= -\rho_3\cot^2\beta +\frac{1}{v^2\sin^2\beta}\left(m_{\eta_H^{}}^2-m_{\eta_A^{}}^2\right).
\end{align}

Therefore, to recap, we are left with 5 new parameters in the active sector ($\tan\beta$, $m_{H^\pm}$, $m_A^{}$, $m_H^{}$ and $M^2$)  and 5 new ones in the inert sector too ($m_{\eta^\pm}$, $m_{\eta_A^{}}$, $m_{\eta_H^{}}$, $\rho_2$ and $\rho_3$) and we will scan over these. Regarding the SM inputs,  
 we use the following values \cite{PDG2014,Koide}:
\begin{align}
&m_Z=91.1876~\text{GeV},~
m_W=80.385~\text{GeV},~
G_F=1.1663787\times 10^{-5}~\text{GeV}^{-2}, \notag\\
&m_t=173.07~\text{GeV},~          
m_b = 3.0~\text{GeV},~
m_c = 0.677~\text{GeV},~V_{cb}= 0.0409,~ V_{ts}= 0.0429, \notag\\
&m_\tau=1.77684~\text{GeV},~
m_\mu=0.105658367~\text{GeV},~~m_h=125~\text{GeV}. 
\end{align}
where $V_{ij}$ are the Cabibbo-Kobayashi-Maskawa matrix elements, and 
the quark masses $m_b$ and $m_c$ are given at the $m_Z^{}$ scale as quoted from Ref.~\cite{Koide}. 

The form factors depend on the three momenta $p_W^{\mu}$, $p_V^{\mu}$ and $q^\mu=p_W^{\mu}+p_V^{\mu}$ for $W$, $V~(=Z,\gamma)$ and $H^\pm$, respectively. 
In the numerical calculation, when $m_{H^\pm}\geq m_W^{}+m_Z^{}$, we take $p_W^2=m_W^2$, $p_Z^2=m_Z^2$ and $q^2=m_{H^\pm}^2$ while 
when $m_{H^\pm}<m_W^{}+m_Z^{}$, we take $p_W^2=(m_{H^\pm}-m_Z)^2$, $p_Z^2=m_Z^2$ and $q^2=m_{H^\pm}^2$ (thereby allowing for below threshold $H^\pm$ decays too). 
For the $H^\pm W^\mp \gamma$ vertex, we take $p_W^2=m_W^2$, $p_\gamma^2=0$ and $q^2=m_{H^\pm}^2$. 

\subsection{Form factors}

\begin{figure}[t]
\begin{center}
\includegraphics[width=70mm]{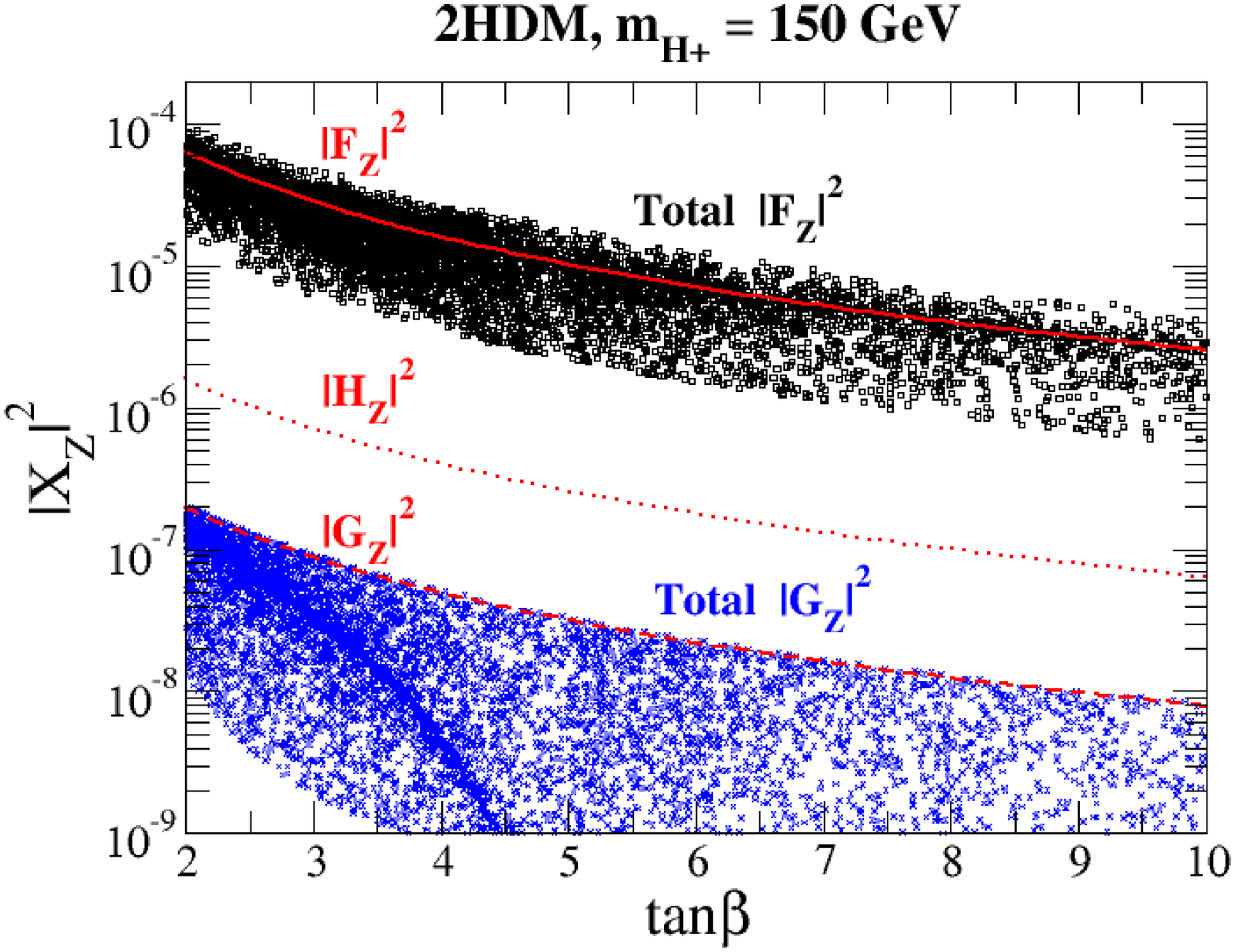} \hspace{5mm}
\includegraphics[width=70mm]{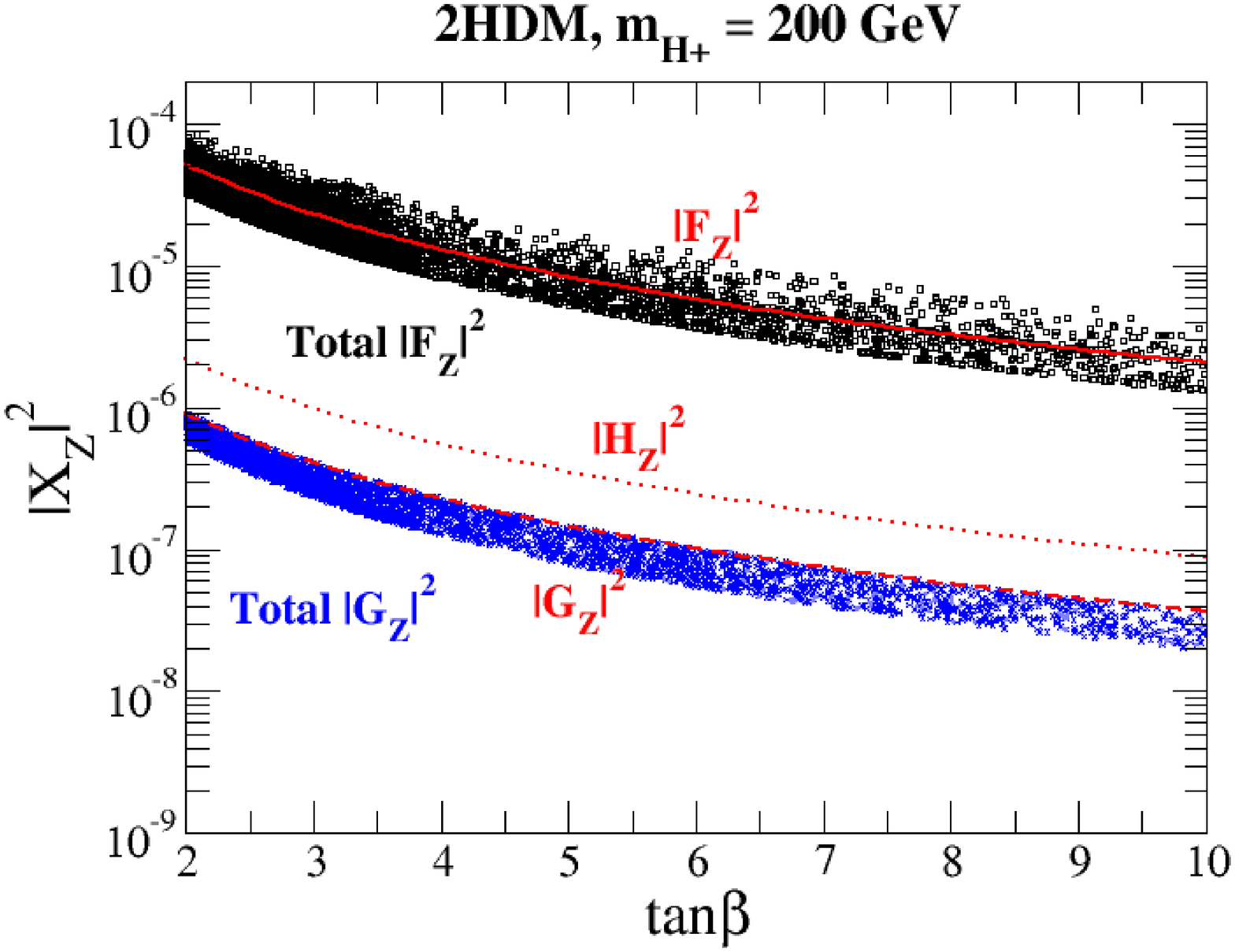}\\ \vspace{2mm}
\includegraphics[width=70mm]{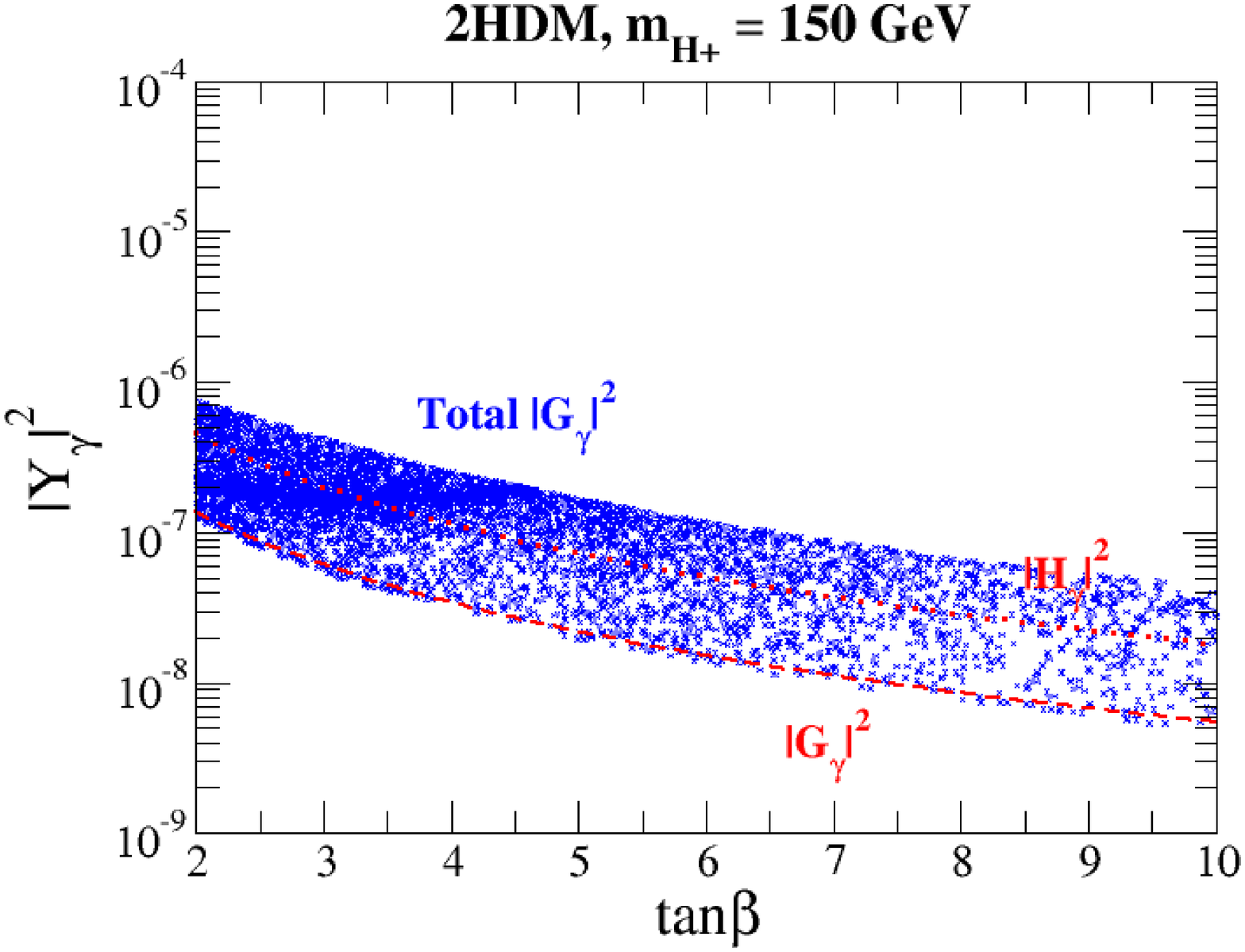} \hspace{5mm}
\includegraphics[width=70mm]{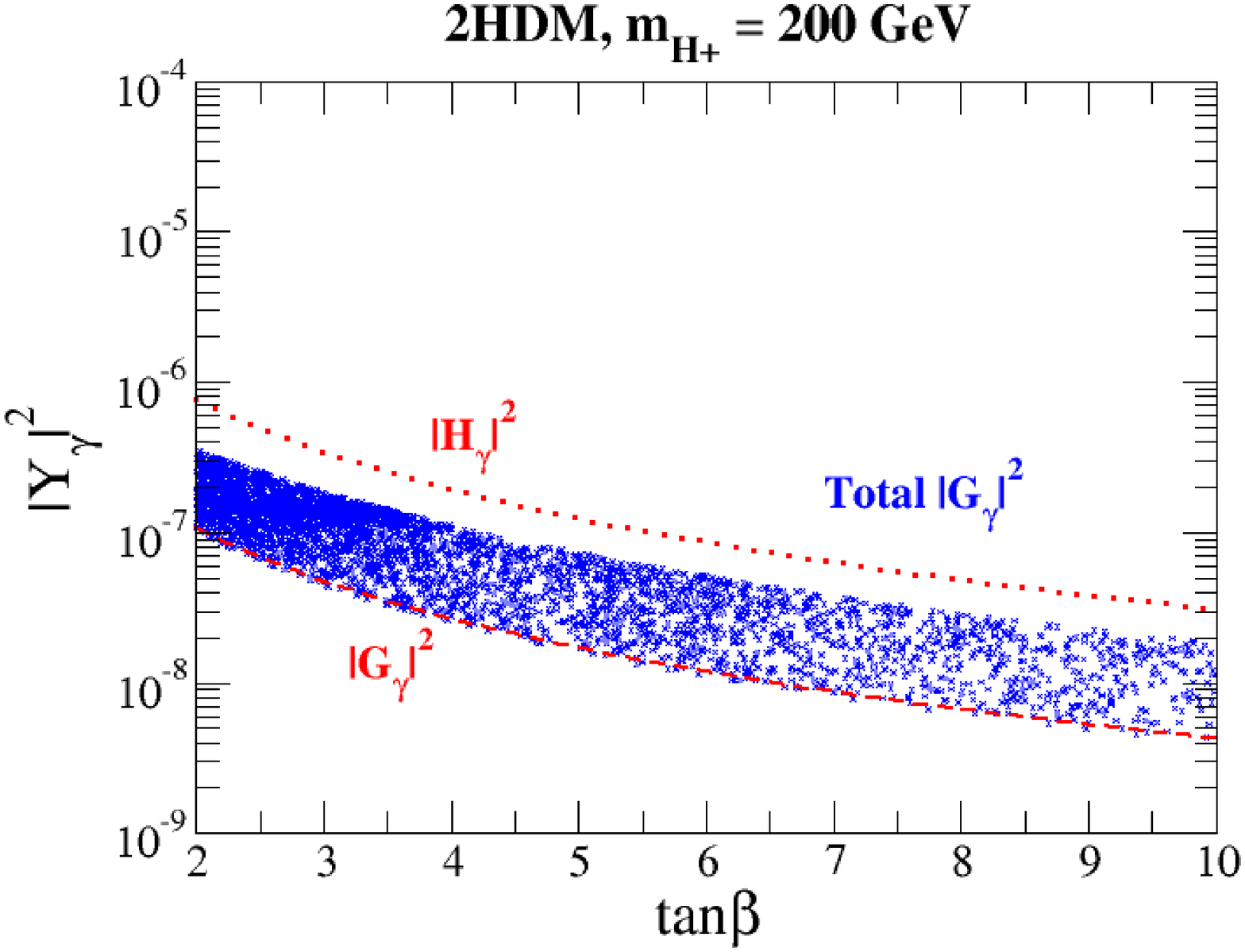} 
\caption{Values of $|X_Z|^2$ ($X=F,~G$ and $H$) (upper panels) and $|Y_\gamma|^2$ ($Y=G$ and $H$) (lower panels) as a function of $\tan\beta$ in the 2HDM
with the Type-I or Type-X Yukawa interactions. 
We take $m_{H^\pm}^{}=150$ GeV (left panels) and 200 GeV (right panels). 
In both the panels, $m_H=m_{H^\pm}$ and $\sin(\beta-\alpha)=1$ are taken. 
The values of $M^2$ and $m_A^2$ are scanned over the ranges of $-400^2<M^2<+400^2$ GeV$^2$ and 
$100 <m_A<260~(350)$ GeV in the left (right) panels, respectively. 
The solid, dashed and dotted (dashed and dotted) curves respectively show the fermion loop contribution to $|F_Z|^2$, $|G_Z|^2$ and $|H_Z|^2$
($|G_\gamma|^2$ and $|H_\gamma|^2$), while
the scatter plots show the total contribution. }
\label{Fsq_2HDM}
\end{center}
\end{figure}

We start by showing the numerical results of the form factors of the $H^\pm W^\mp Z$ and $H^\pm W^\mp \gamma$ vertices. 
In order to see how the inert scalar boson loops can change the prediction, 
we first show the result in the 2HDM under the constraints from 
unitarity, vacuum stability and the EW parameters as discussed in Sec.~IV.  Then, we move on to the 3HDM.

In Fig.~\ref{Fsq_2HDM}, the values of $|X_Z|^2$ ($X=F,~G$ and $H$) and $|Y_\gamma|^2$ $(Y=G$ and $H$)
are respectively plotted in the upper and lower panels as a function of $\tan\beta$ in the case of $\sin(\beta-\alpha)=1$ and $m_H^{}=m_{H^\pm}$. 
The left (right) panel shows the case of $m_{H^\pm}=150~(200)$ GeV. 
The solid, dashed and dotted (dashed and dotted) curves respectively show the fermion loop contribution to $|F_Z|^2$, $|G_Z|^2$ and $|H_Z|^2$ ($|G_\gamma|^2$ and $|H_\gamma|^2$)
whereas the black and blue (blue) scatter plots are the total contribution to $|F_Z|^2$ and $|G_Z|^2$ ($|G_\gamma|^2$), respectively. 
For the boson loop contribution, we scan the parameters over the intervals $-400^2$  GeV$^2<M^2<400^2$ GeV$^2$ and 
$100$ GeV $ <m_A<260~(350)$ GeV in the left (right) panels. 
We note that $m_A^{}\gtrsim 260~(350)$ GeV when $m_{H^\pm}(=m_H^{})=150$ (200) GeV 
is excluded by the constraint from the $S$ parameter at 95\% CL. 
We also note that only the fermion loop contributes to $H_Z$ and $H_\gamma$. 

\begin{figure}[t]
\begin{center}
\includegraphics[width=70mm]{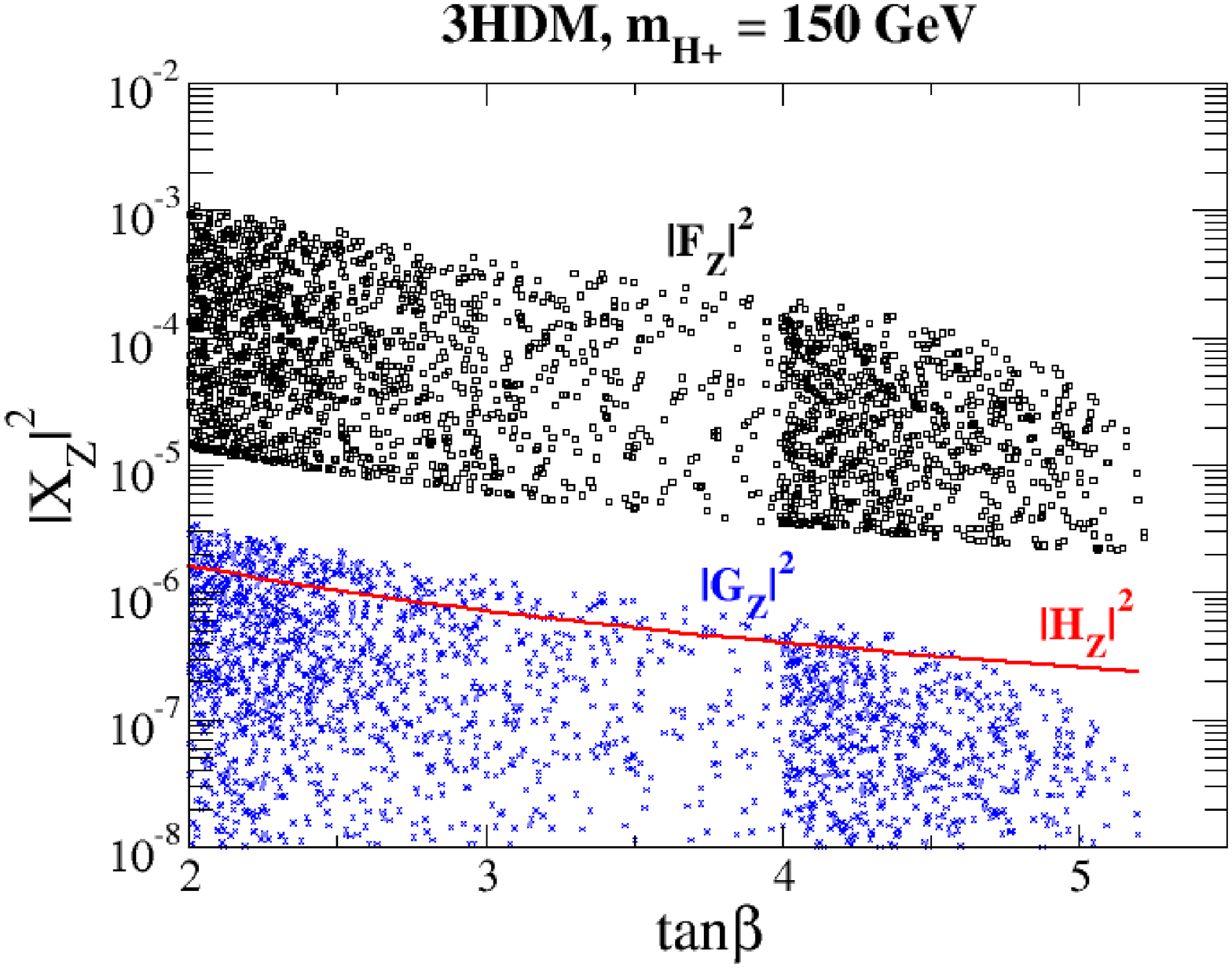}\hspace{5mm}
\includegraphics[width=70mm]{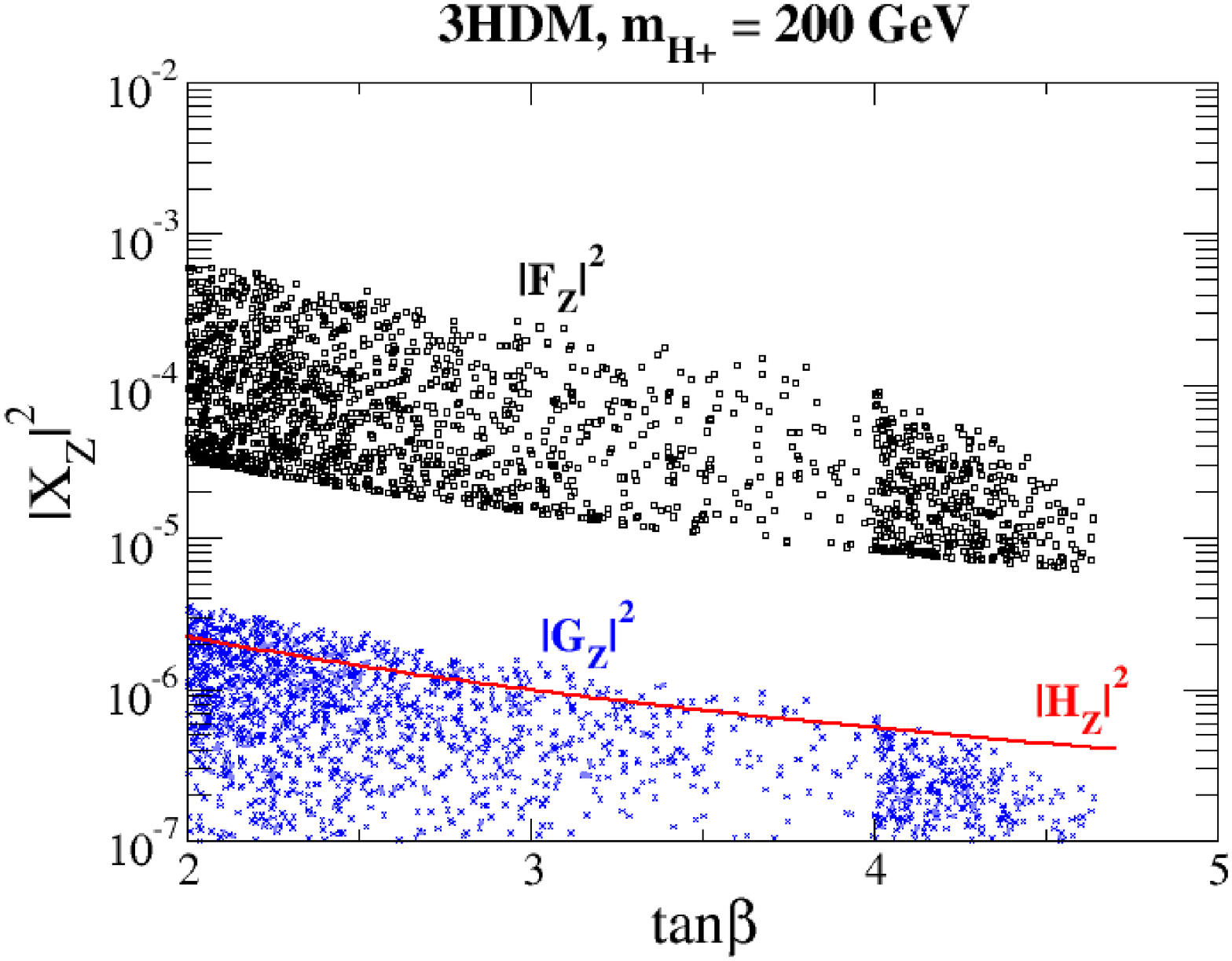} 
\\ \hspace{5mm}
\includegraphics[width=70mm]{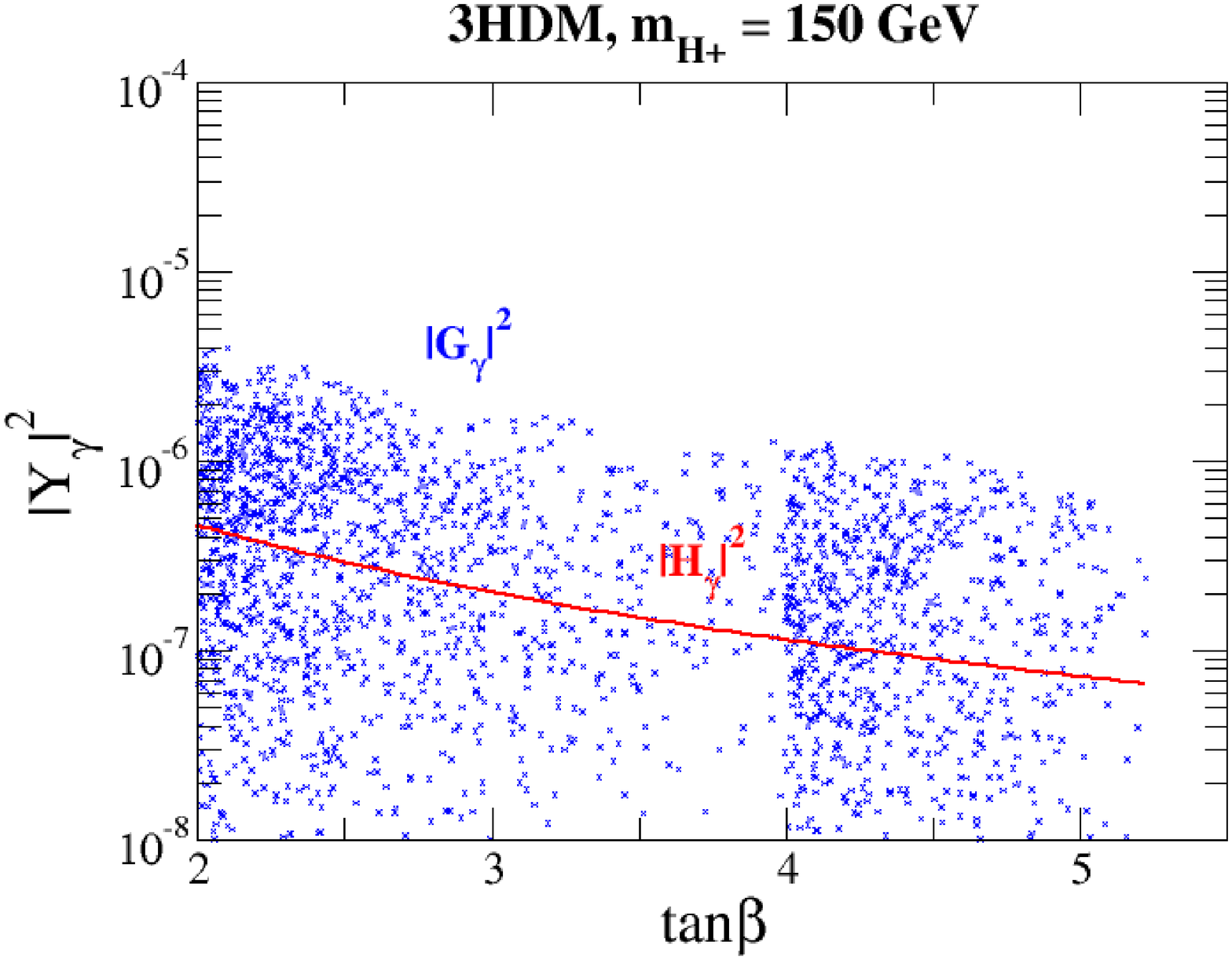}\hspace{5mm}
\includegraphics[width=70mm]{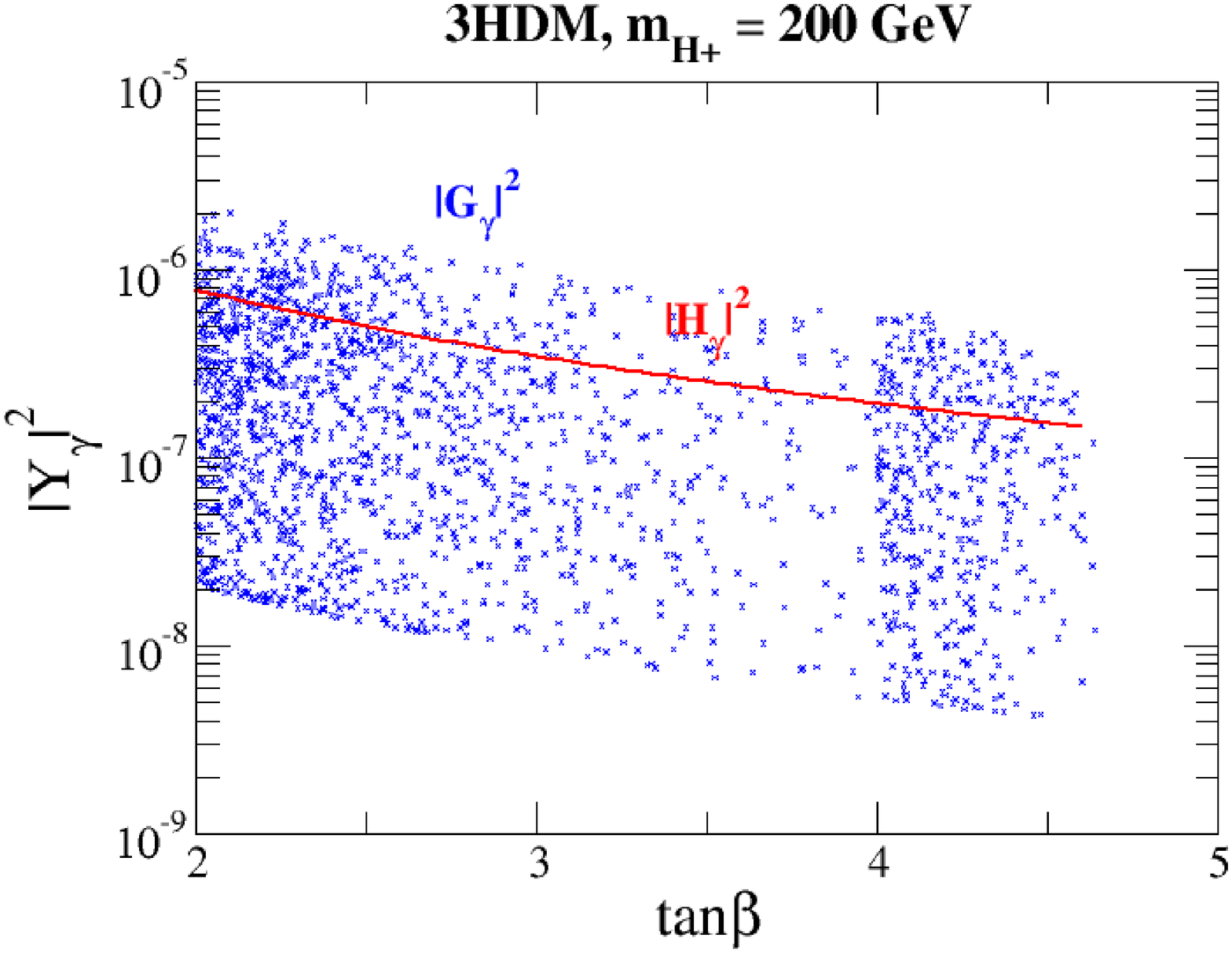} 
\caption{Values of $|X_Z|^2$ ($X=F,~G$ and $H$) (upper panels) and $|Y_\gamma|^2$ ($Y=G$ and $H$) (lower panels) as a function of $\tan\beta$ in the 3HDM 
with $m_{\eta_A}^{}=400$ GeV.  
We take $m_{H^\pm}^{}=150$ GeV (left panels) and 200 GeV (right panels). 
All the other parameters are taken as given in Eq. (\ref{parameter}). 
In the upper panel, the black scatter plot shows the values of $|F_Z|^2$.  
In all the panels, the blue scatter plot and the solid curve respectively represent $|G_V|^2$ and $|H_V|^2$ ($V=Z,\gamma$). 
}
\label{Fsq_3HDM1}
\end{center}
\end{figure}

We can see that the value of $|F_Z|^2$ is the biggest of all the form factors as we expected in Sec.~III, because of the $m_t^2$ dependence. 
Typically, $|F_Z|^2$ is more than one order of magnitude larger than $|G_Z|^2$ and $|H_Z|^2$. 
In addition, all the squared form factors decrease as $\tan\beta$ is getting larger, because the top Yukawa coupling is proportional to $\cot\beta$.  
The maximal allowed value of $|F_Z|^2$ is obtained to be about $10^{-4}$ at $\tan\beta\simeq 2.5$ in both the cases of $m_{H^\pm}=150$ GeV and 200 GeV.  
For the $H^\pm W^\mp\gamma$ vertex, the maximal allowed values of $|G_\gamma|^2$ and $|H_\gamma|^2$ are order of $10^{-6}$ at $\tan\beta\simeq 2$.

\begin{figure}[t]
\begin{center}
\includegraphics[width=70mm]{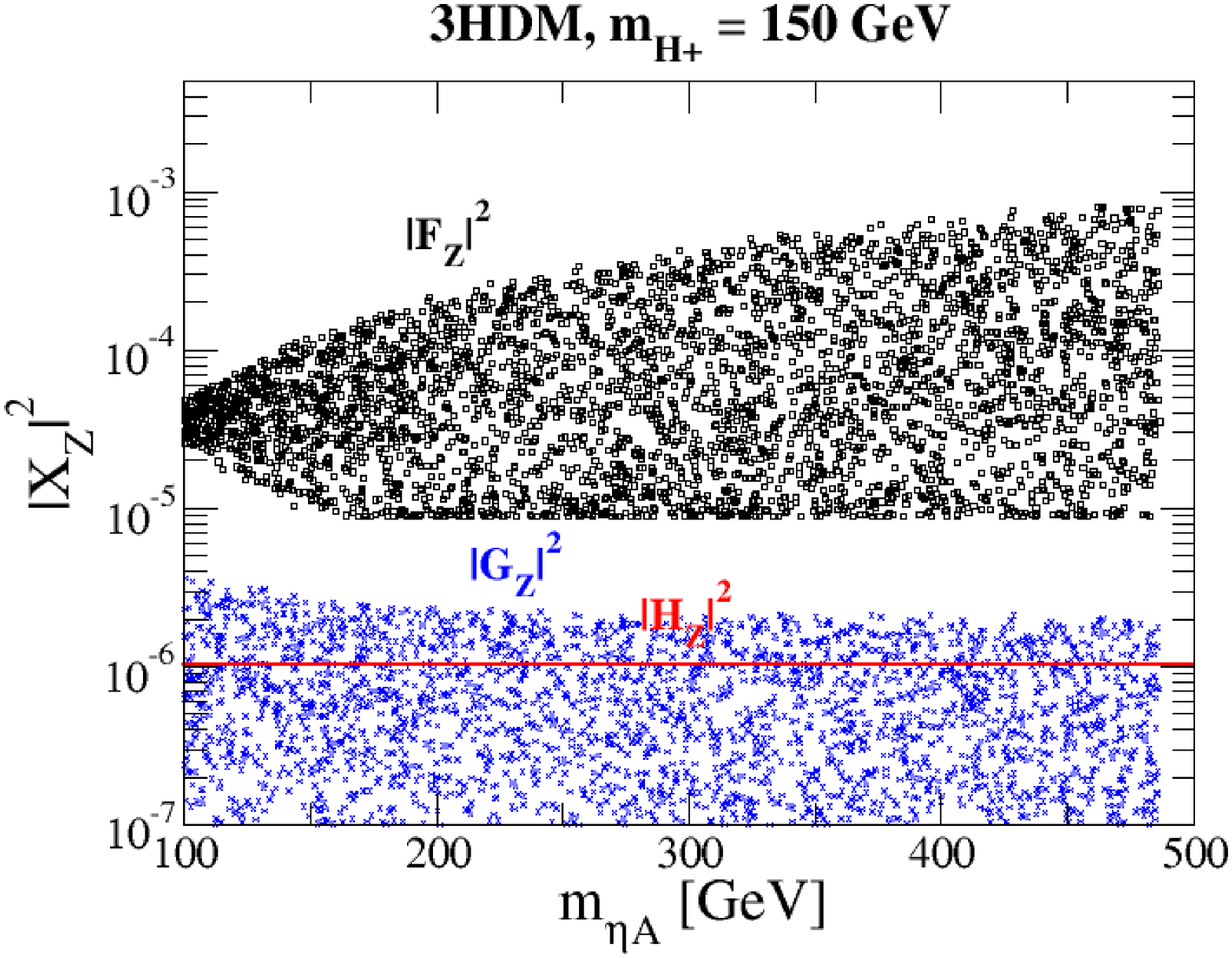}\hspace{5mm}
\includegraphics[width=70mm]{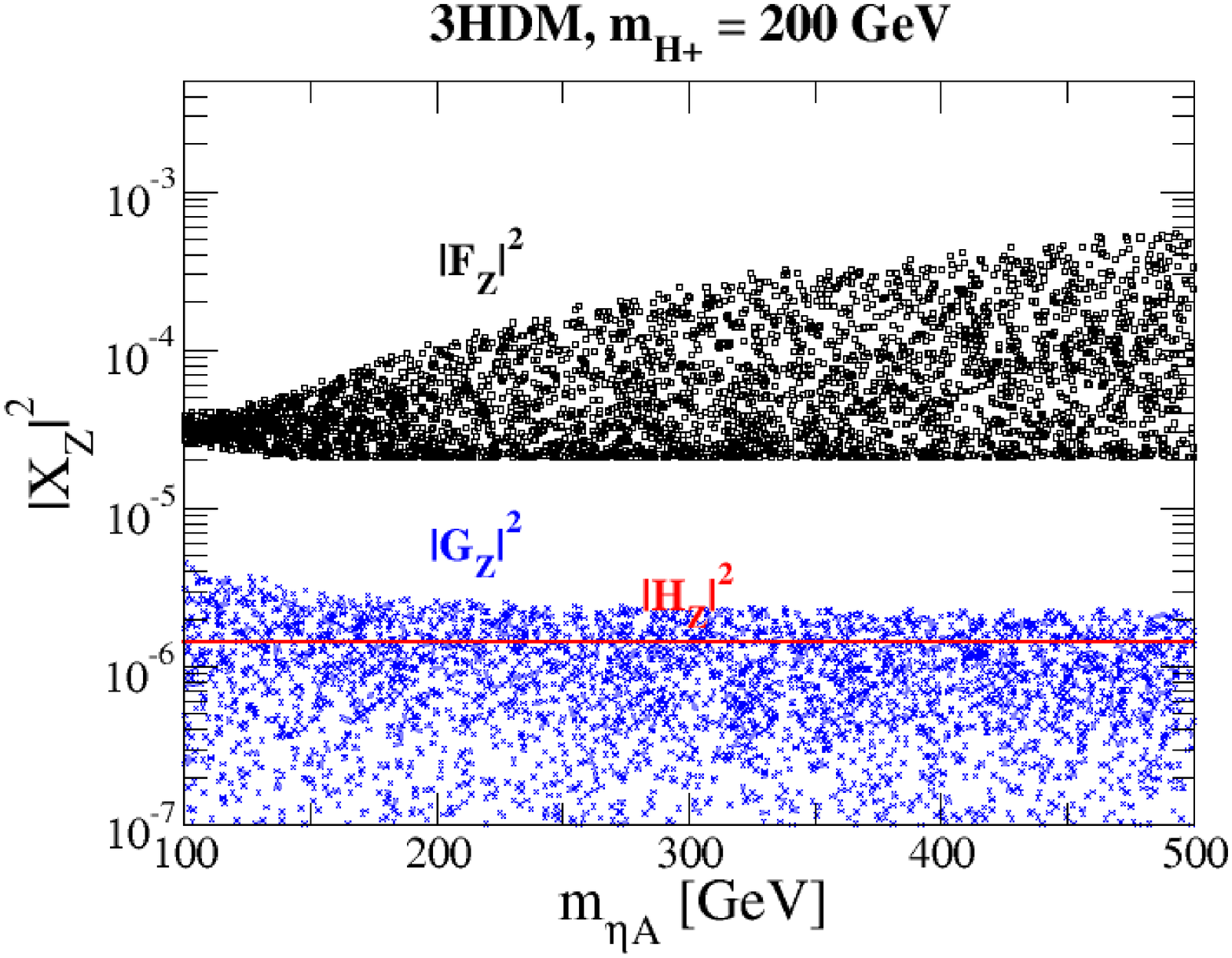}
\\ \hspace{5mm}
\includegraphics[width=70mm]{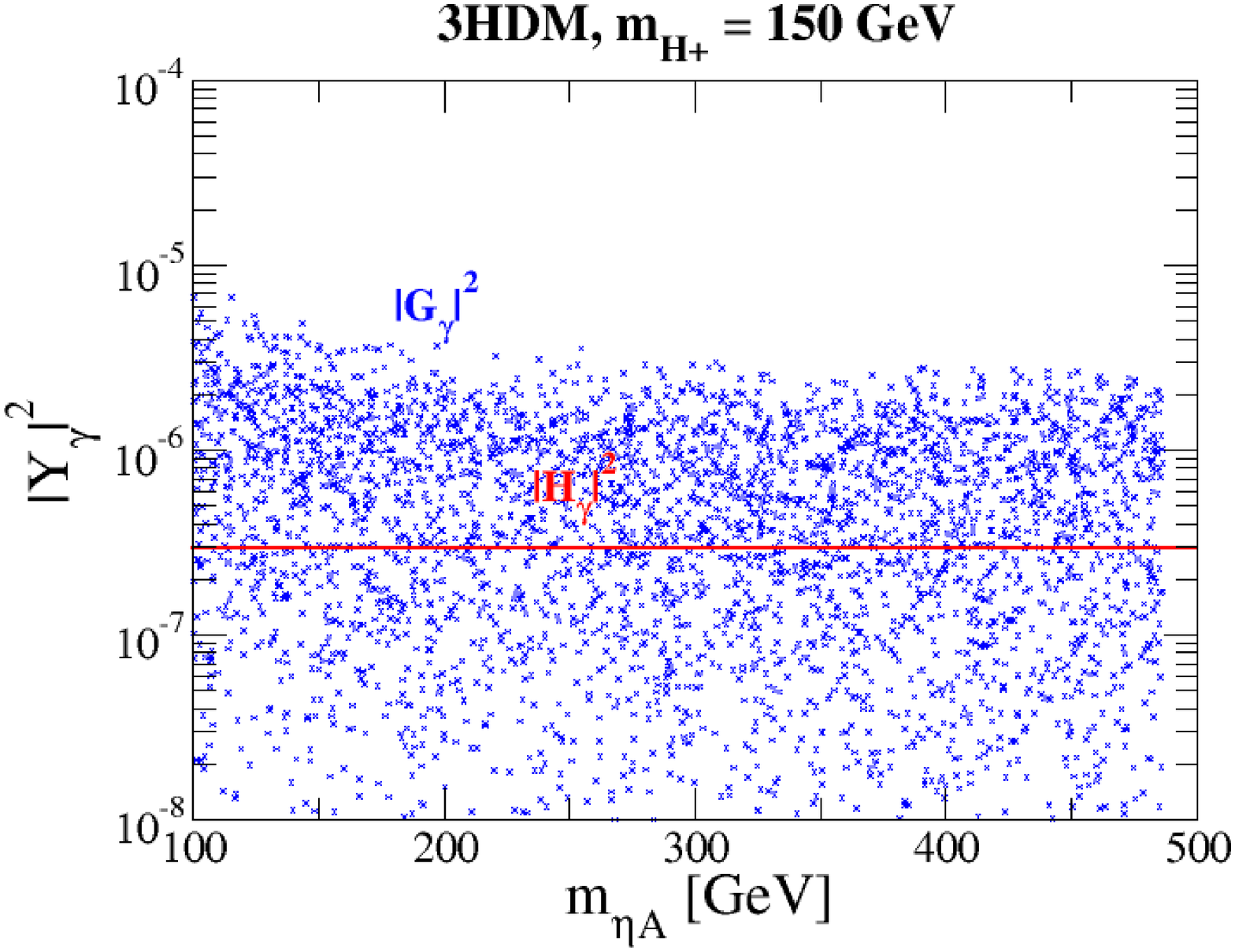}\hspace{5mm}
\includegraphics[width=70mm]{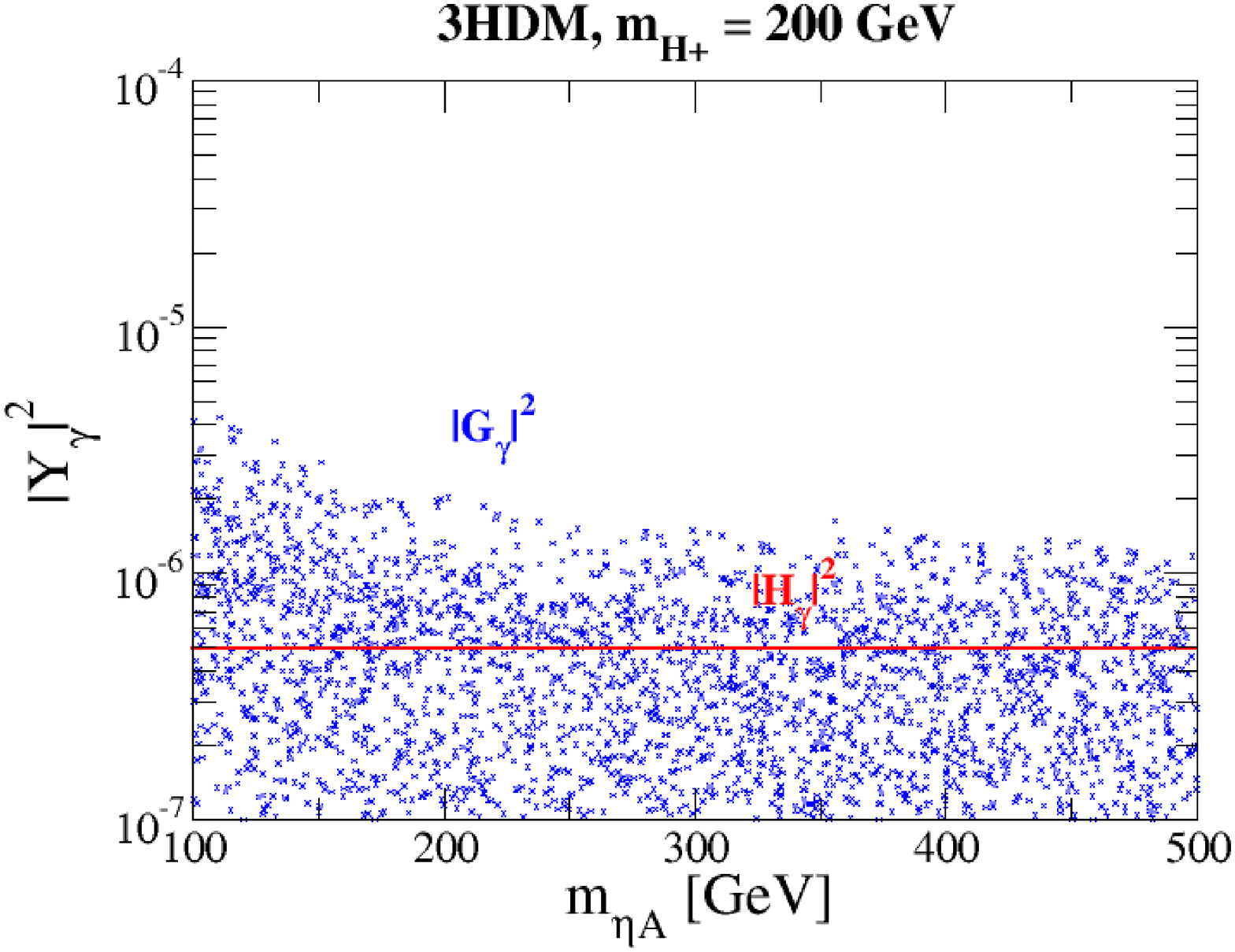} 
\caption{
Values of $|X_Z|^2$ ($X=F,~G$ and $H$) (upper panels) and $|Y_\gamma|^2$ ($Y=G$ and $H$) (lower panels) as a function of $m_{\eta_A^{}}$ in the 3HDM 
with $\tan\beta=2.5$.  
We take $m_{H^\pm}^{}=150$ GeV (left panels) and 200 GeV (right panels). 
All the other parameters are taken as given in Eq. (\ref{parameter}). 
In the upper panel, the black scatter plot shows the values of $|F_Z|^2$.  
In all the panels, the blue scatter plot and the solid curve respectively represent $|G_V|^2$ and $|H_V|^2$ ($V=Z,\gamma$). 
}
\label{Fsq_3HDM2}
\end{center}
\end{figure}

Regarding the 3HDM, 
as we see from Eq.~(\ref{fz_app}), $F_Z$ is logarithmically enhanced by $m_{\eta_A^{}}$ in the case of $m_{\eta^\pm}=m_{\eta_H^{}}$. 
However, a too large mass difference between $\eta_A^{}$ and $\eta^\pm$ is excluded by the $S$ parameter as shown in Fig.~\ref{stu_fig}
in the case of $m_{H^\pm}=m_A^{}=m_{H}^{}$ or $\Delta T = 0$. 
We thus take a mass difference between $H^\pm$ and $A/H$ with $m_H^{}=m_A^{}$ to 
avoid the constraint by the effect of non-zero $\Delta T$. 
From the above reason, we consider the following parameter conditions in the forthcoming calculations: 
\begin{align}
&m_A^{}=m_H^{}=m_{H^\pm} + 50~\text{GeV},\quad M^2=m_{H^\pm}^2, \notag\\  
&m_{\eta^\pm}=m_{\eta_H^{}}=\frac{1}{2}m_A^{},\quad m_{\eta_A^{}}>m_{\eta^\pm},\quad -10<\rho_2,\rho_3<10. \label{parameter}
\end{align}
We note that, 
in this setup, $\eta_H^{}$ corresponds to the DM candidate. 
The measured relic abundance of  DM\footnote{Because the DM phenomenology is not the main topic of this 
paper, we do not perform the detailed analysis such as the calculation of the  (co)annihilation cross sections of the 
DM candidate. } can be satisfied  
by the resonant process of $\eta_H^{} \eta_H^{} \to A/H \to f\bar{f}$. 

In Fig.~\ref{Fsq_3HDM1}, 
the values of $|X_Z|^2$ ($X=F,~G$ and $H$) and $|Y_\gamma|^2$ $(Y=G$ and $H$)
are respectively shown in the upper and lower panels as a function of $\tan\beta$  with $m_{\eta_A^{}}=400$ GeV. 
The left (right) panel shows the case of $m_{H^\pm}=150~(200)$ GeV. 
In the upper panel, the black scatter plots show the values of $|F_Z|^2$.  
In all the panels, the blue scatter plot and the solid curve respectively represent $|G_V|^2$ and $|H_V|^2$ ($V=Z,\gamma$). 
Similar to the results in the 2HDM, $|F_Z|^2$ is the biggest of all the squared form factors also in the 3HDM, and 
all the squared form factors become smaller when $\tan\beta$ becomes large.  Remarkably,
at $\tan\beta= 2$, we obtain $|F_Z|^2\simeq 10^{-3}$, which is one order of magnitude larger than $|F_Z|^2$ in the 2HDM. 

In Fig.~\ref{Fsq_3HDM2}, we show the $m_{\eta_A^{}}$ dependence of the squared form factors in the case of $\tan\beta=2.5$. 
We take $m_{H^\pm}=150~(200)$ GeV in the left (right) panel. 
The description of the objects in the figure is the same as in Fig.~\ref{Fsq_3HDM1}. 
Clearly, we can see that only $|F_Z|^2$ is enhanced as $m_{\eta_A^{}}$ is getting larger. 
The maximal allowed value of $|F_Z|^2$ is about $10^{-3}$ at $m_{\eta_A^{}}\simeq 500$ GeV. 

\subsection{Branching fractions of $H^\pm$}

Next, we discuss the decay branching ratios of $H^\pm$. 
As we see in Figs.~\ref{Fsq_3HDM1} and \ref{Fsq_3HDM2} that 
the form factor $F_Z$ is much larger than $G_Z$ and $H_Z$, we only keep the term proportional to $|F_Z|^2$ for the $H^\pm \to WZ$ decay. 
When $m_{H^\pm}^{}> m_W+m_Z$, the on-shell decay of $H^\pm \to W^\pm Z$ opens and its decay rate is calculated as
\begin{align}
\Gamma(H^\pm \to W^\pm Z) &= \frac{\sqrt{2}G_F}{16\pi}m_{H^\pm}^3 \lambda^{1/2}(x_W^{},x_Z^{})
c_W^2[\lambda(x_W^{},x_Z^{})+12x_W^{}x_Z^{}]|F_Z|^2,\label{Gam_HWZ}
\end{align}
where $x_W^{}=m_W^2/m_{H^\pm}^2$ and $x_Z^{}=m_Z^2/m_{H^\pm}^2$. 
If $m_{H^{\pm}}^{}$ is smaller than $m_W^{}+m_Z^{}$, the off-shell decay modes $H^\pm \to W^\pm Z^*$ and 
$H^\pm \to W^{\pm *} Z$ are allowed. 
The decay rate with three body final states is given by 
\begin{align}
&\sum_{f,f'}\Gamma(H^\pm \to W^{\pm *} Z \to Zf\bar{f}') = \frac{9g^4m_W^2}{256\pi^3 m_{H^\pm}}|F_Z|^2 F_3\left(x_Z^{},x_W^{}\right), \\
&\sum_{f}\Gamma(H^\pm \to W^\pm Z^* \to Wf\bar{f}) = \frac{3g^4m_Z^2}{512\pi^3 m_{H^\pm}}|F_Z|^2 
\left(7-\frac{40}{3}s_W^2 + \frac{160}{9}s_W^4\right)F_3\left(x_W^{},x_Z^{}\right), 
\end{align}
where 
\begin{align}
&F_3(x,y^*)=\frac{\arctan\left[\frac{(1-x)\sqrt{-\lambda(x,y^*)}}{y^*(1+x)-(1-x)^2}\right]
+\pi}{4x\sqrt{-\lambda(x,y^*)}}
 \Big[(1-y^*)^3-3x^3+(9y^*+7)x^2-5(1-y^*)^2x\Big] \notag\\
&+\frac{1}{24xy^*}\Big\{(x-1)[6y^{*2}+y^*(39x-9)+2(1-x)^2]-3y^*[y^{*2}+2y^*(3x-1)-x(3x+4)+1]\ln x\Big\}. 
\end{align}
We note that the argument $y^*$ is for the ratio of squared masses of a virtual gauge boson to that of $H^\pm$, e.g., 
for the $H^\pm \to W^{\pm *} Z$ case, we should use $F_3(m_Z^2/m_{H^\pm}^2,m_W^2/m_{H^\pm}^2)$. 
The decay rate for $H^\pm\to W^\pm \gamma$ is given by 
\begin{align}
\Gamma(H^\pm \to W^\pm \gamma) &= \frac{\sqrt{2}G_F}{8\pi}m_{H^\pm}^3(1-x_W^{})^3\left(|G_\gamma|^2+|H_\gamma|^2\right).  \label{Gam_HWG}
\end{align}

\begin{figure}[t]
\begin{center}
\includegraphics[width=50mm]{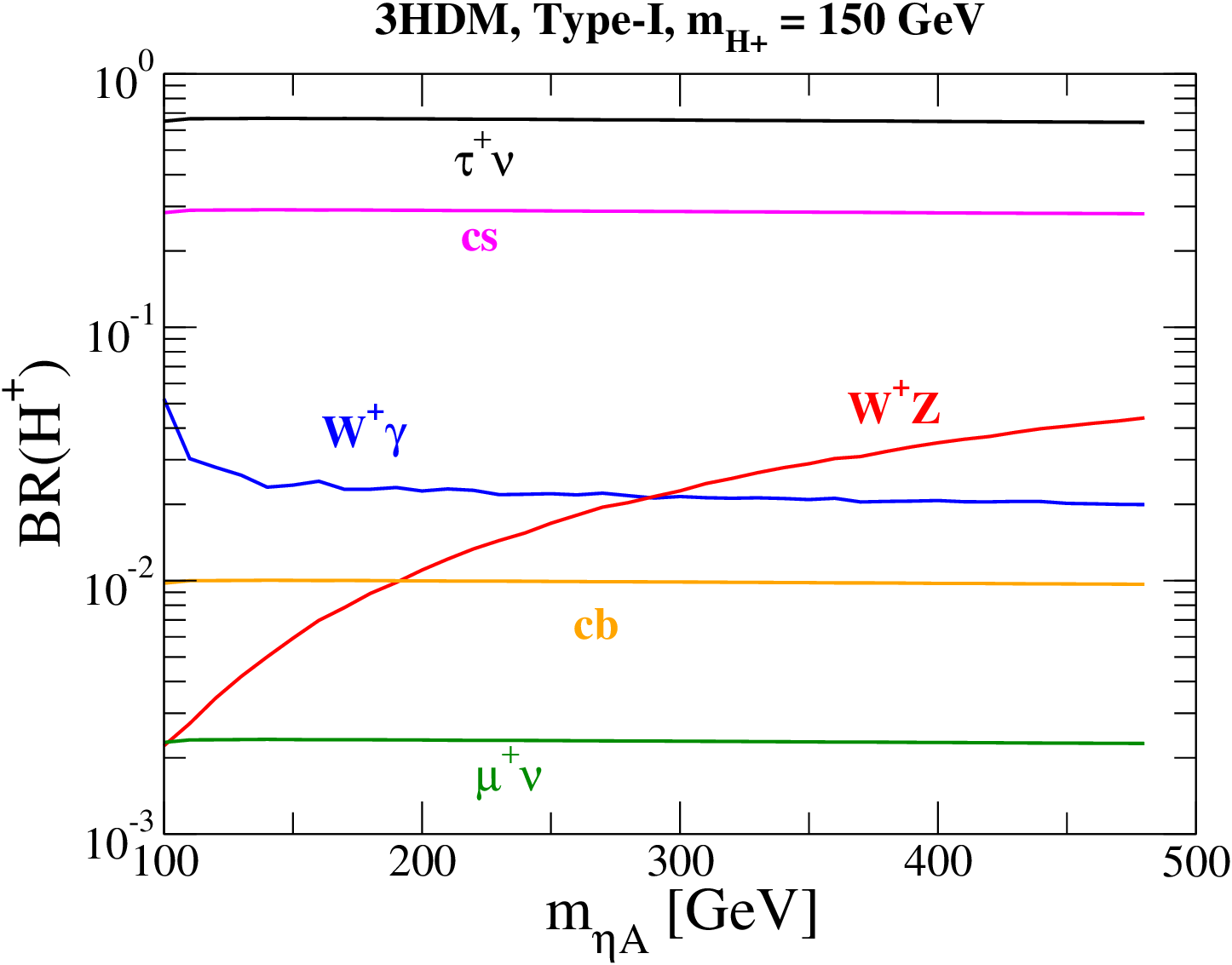} \hspace{5mm}
\includegraphics[width=50mm]{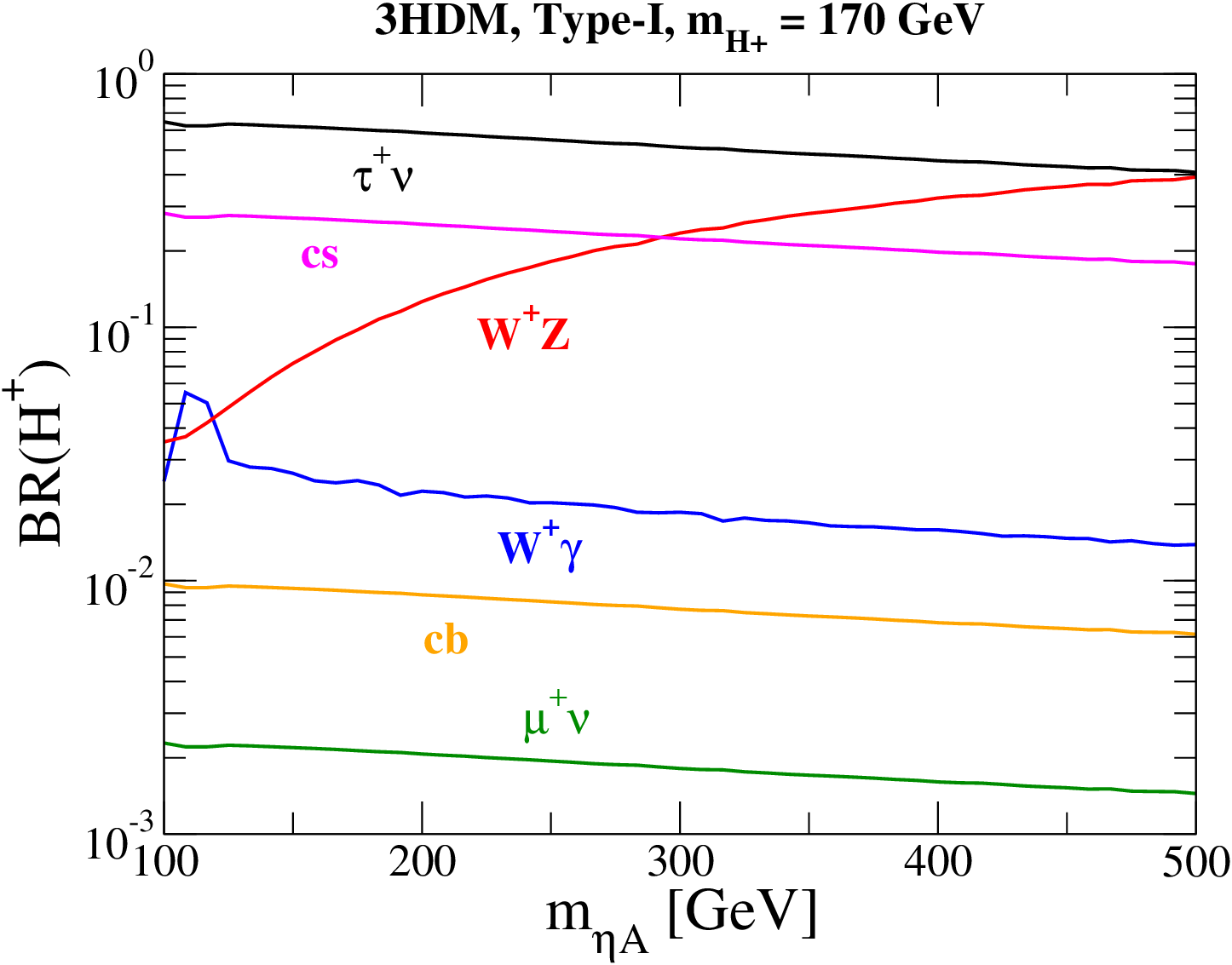} \hspace{5mm}
\includegraphics[width=50mm]{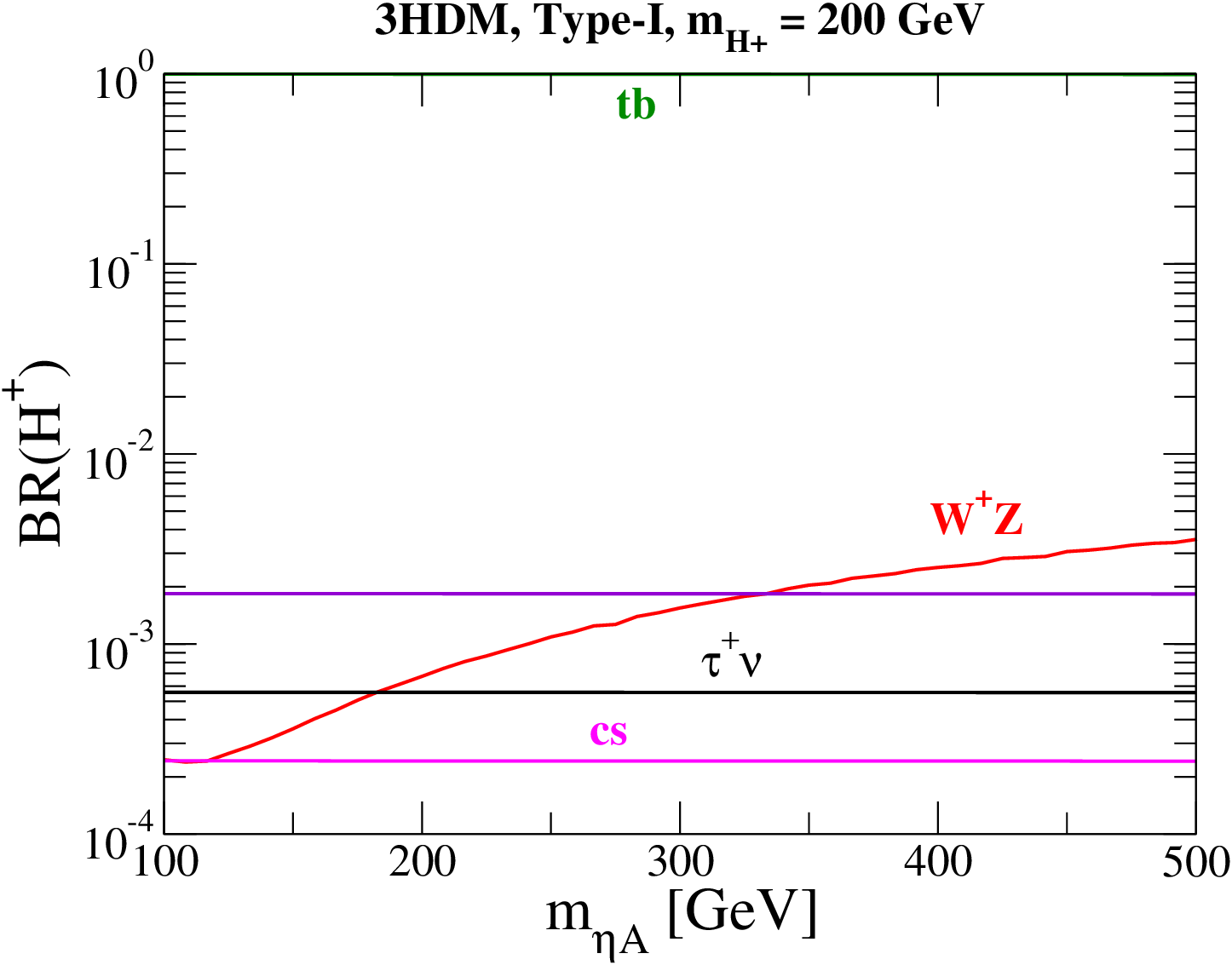}
\caption{Branching fractions of $H^\pm$ as a function of $m_{\eta_A}$  in the Type-I Yukawa interaction with $\tan\beta=2.5$. 
We take $m_{H^\pm}=150$ GeV (left), 170 GeV (center) and 200 GeV (right).   }
\label{BR1}
\vspace{1cm}
\includegraphics[width=50mm]{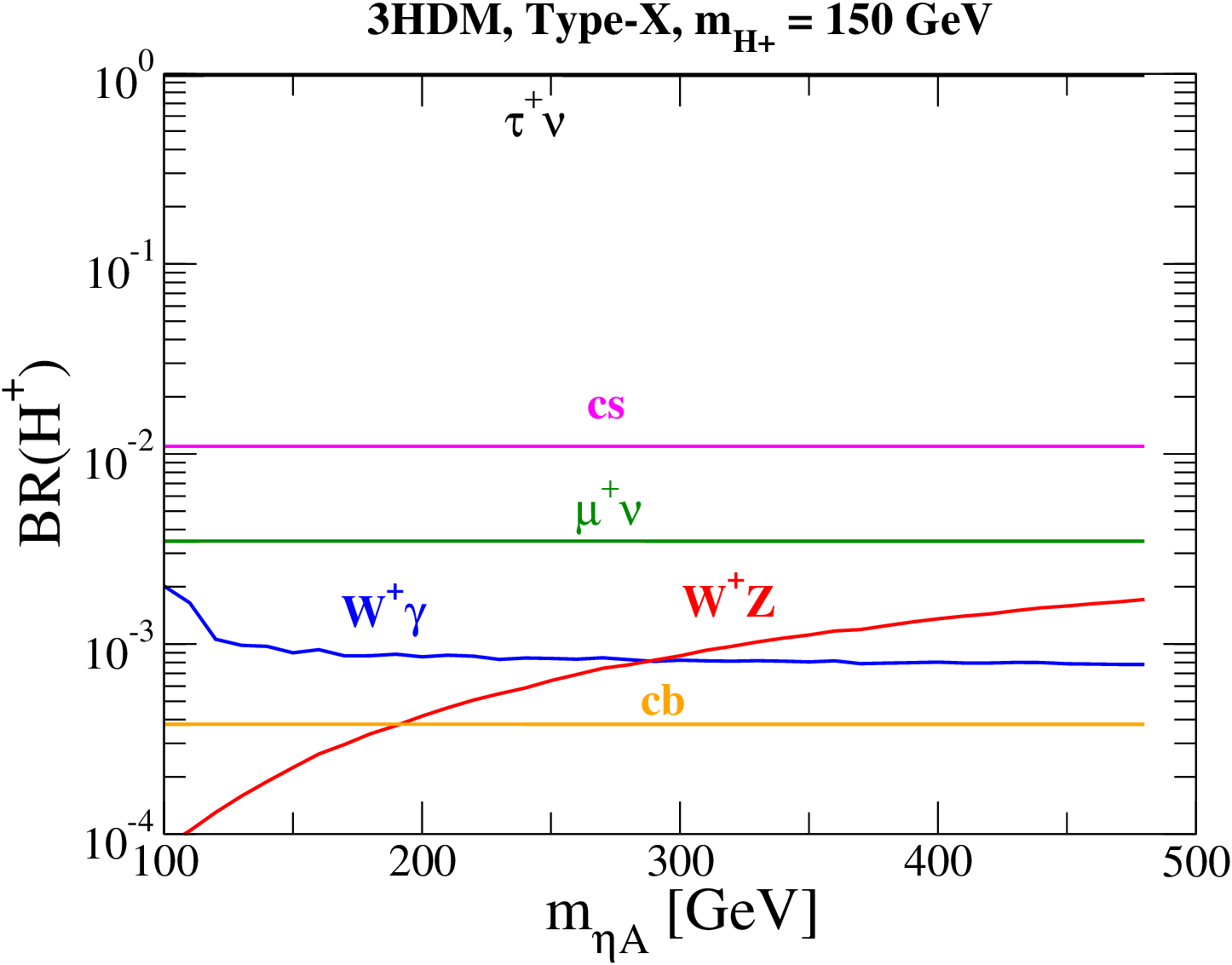} \hspace{5mm}
\includegraphics[width=50mm]{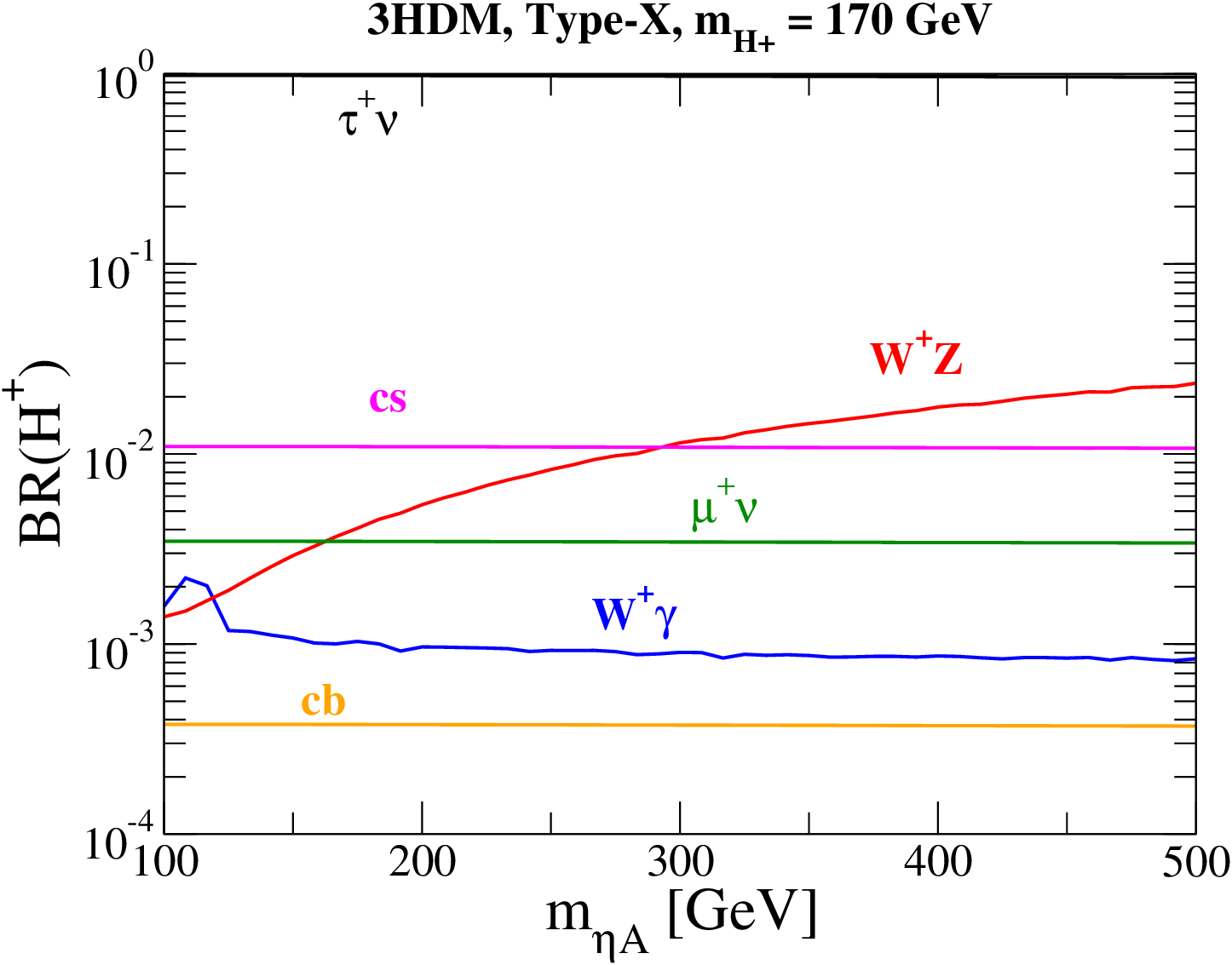}\hspace{5mm}
\includegraphics[width=50mm]{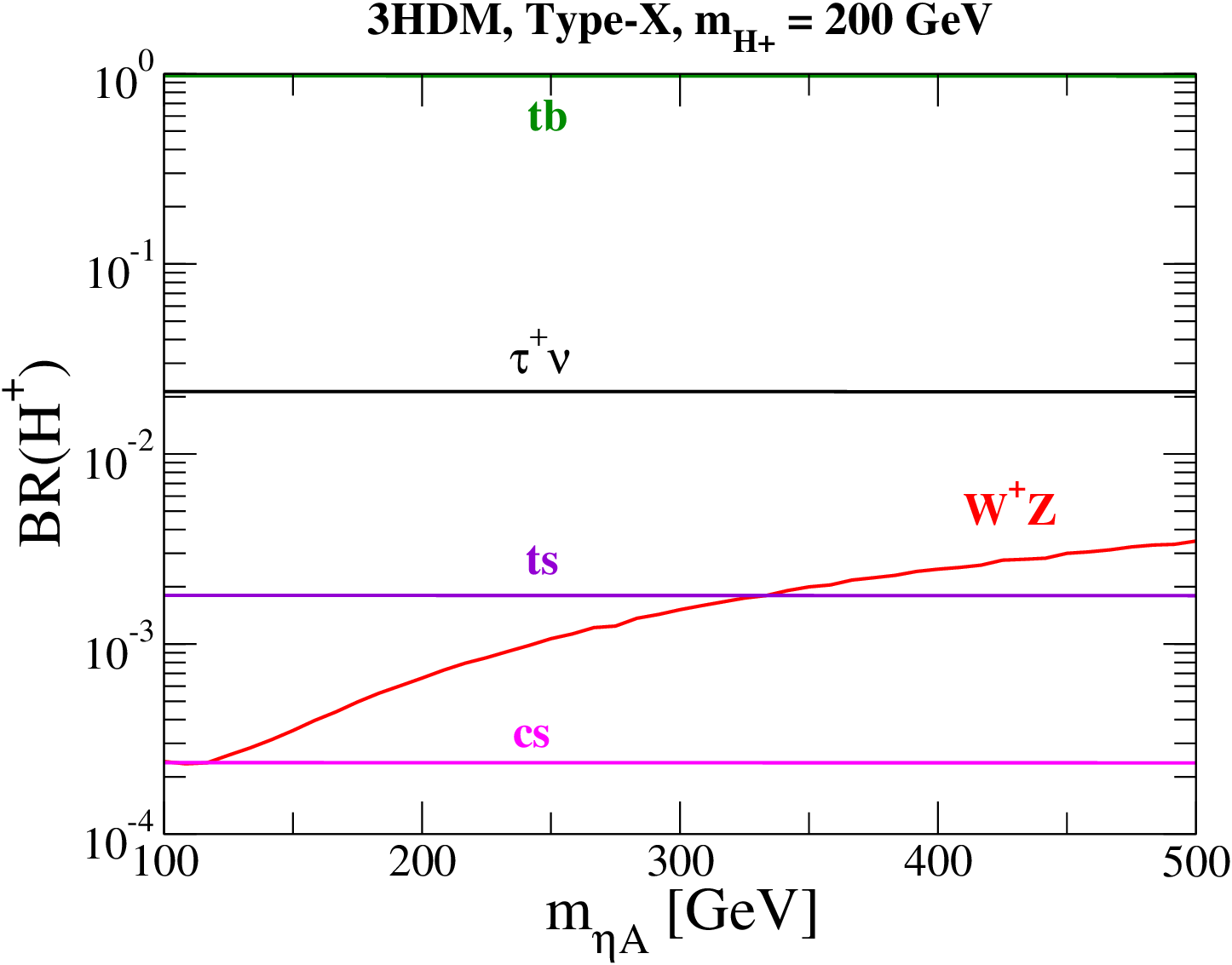}
\caption{Branching fractions of $H^\pm$ as a function of $m_{\eta_A}$  in the Type-X Yukawa interaction with $\tan\beta=2.5$. 
We take $m_{H^\pm}=150$ GeV (left), 170 GeV (center) and 200 GeV (right). 
  }
\label{BRX}
\end{center}
\end{figure}

In Fig.~\ref{BR1}, we show the branching fractions of $H^\pm$ as a function of $m_{\eta_A^{}}^{}$ in the 3HDM with the Type-I Yukawa interaction. 
We take $m_{H^\pm}=150$ (left), 170 (center) and 200 GeV (right). 
The value of $\tan\beta$ is fixed to be 2.5 in all the panels. 
In these plots, we scan the values of $\rho_2$ and $\rho_3$ in the range of $-10$ to $+10$ and extract the set of $(\rho_2,\rho_3)$ combinations
 giving the maximal value of the decay rate $\Gamma(H^\pm \to WZ)$. 
Further, for the case of $m_{H^\pm}<m_W+m_Z$, we show the branching fraction of 
$H^\pm \to W^\pm Z$ as the sum of the branching fractions of $H^\pm \to W^\pm Z^*$ and $H^\pm \to W^{\pm *} Z$. 
In all the plots, the behavior of $m_{\eta_A^{}}$ in the $H^\pm \to W^\pm Z$ decay is similar to that of $|F_Z|^2$ shown in Fig.~\ref{Fsq_3HDM2}.  
In the case of $m_{H^\pm}=150$ GeV, 
although BR($H^\pm\to W^\pm Z$) benefits from the enhancement of $|F_Z|^2$, 
its rate is smaller than BR($H^\pm\to W^\pm \gamma$) when $m_{\eta_A^{}}\lesssim 300$ GeV. 
This can be understood by the suppression of the decay rate of $H^\pm\to W^\pm Z$ due to the off-shell effect of the $W^\pm$ or $Z$ bosons.
Therefore, we obtain a larger value of BR($H^\pm\to W^\pm Z$) in the case of $m_{H^\pm}=170$ GeV because of the smaller off-shell effect. 
However, once $m_{H^\pm}$ exceeds the top quark mass, both the branching fractions of $H^\pm \to W^\pm Z$ and 
$H^\pm \to W^\pm \gamma$ are significantly suppressed by the $H^\pm \to tb$ decay. 
We find that the maximal value of BR($H^\pm \to W^\pm Z$) is about 4\%, 40\% and 0.4\% in the cases of $m_{H^\pm}=150$, 170 and 200 GeV, respectively.  

In Fig.~\ref{BRX}, we also show the branching fraction of $H^\pm$ in the Type-X Yukawa interaction with $\tan\beta=2.5$. 
Although we observe a similar behavior of BR($H^\pm\to W^\pm Z$) and BR($H^\pm\to W^\pm \gamma$) as seen in Fig.~\ref{BR1}, 
their maximal values are smaller than those in the case of the Type-I Yukawa interaction. 
This is because in the Type-X Yukawa interaction,   
the decay rate of the $H^\pm \to \tau^\pm \nu$ mode is enhanced by $\tan^2\beta$. 
Here, the maximal value of BR($H^\pm \to W^\pm Z$) is about 0.2\%, 2\% and 0.3\% in the cases of $m_{H^\pm}=150$, 170 and 200 GeV, respectively.

\subsection{Cross sections at the LHC}

\begin{table}[t]
\begin{center}
{\renewcommand\arraystretch{1.2}
\begin{tabular}{l|lllll}\hline\hline
                  &Type-I &Type-X  \\  \hline
Br($t\to H^\pm b$) [\%]&( 3.3, 1.10, 4.7$\times 10^{-3}$)  & (3.3, 1.1, 4.7$\times 10^{-3}$) \\  \hline
Br$(H^\pm \to W^\pm Z)$ [\%]&(0.66, 3.5, 33) &  (0.025, 0.14, 1.8 )   \\  \hline        
Br$(H^\pm \to W^\pm \gamma)$ [\%]&(1.6, 2.1, 1.6) & (0.059, 0.081, 0.087)   \\  \hline     
$\sigma_{S,Z}^{\text{top}} $ [fb] &(390, 700, 29) &(15, 28, 1.6) \\  \hline
$\sigma_{S,\gamma}^{\text{top}} $ [fb] &(940, 420, 1.4) &(35, 16, 0.075) \\  \hline
$\sigma_{S,Z}^{\text{EW}} $ [fb] &(2.3, 7.5, 46) &(0.087, 0.30, 2.5) \\  \hline
$\sigma_{S,\gamma}^{\text{EW}} $ [fb] &(5.5, 4.5, 2.2) &(0.20, 0.17, 0.12) \\  \hline\hline
\end{tabular}}
\caption{The branching fractions and the cross sections in the 3HDM with a Type-I and Type-X Yukawa interaction. 
We take $\tan\beta=2.5$ and $m_{\eta_A^{}}=400$ GeV. 
The numbers in the bracket correspond to the result of $m_{H^\pm}^{}=$130, 150 and 170 GeV from left to right. }
\label{res1}
\vspace{1cm}
{\renewcommand\arraystretch{1.2}
\begin{tabular}{l|lllll}\hline\hline
                  & Type-I & Type-X\\  \hline
Br($t\to H^\pm b$) [\%]&(1.3, 0.43, 1.8$\times 10^{-3}$) & (1.3, 0.43, 1.8$\times 10^{-3}$) \\  \hline
Br$(H^\pm \to W^\pm Z)$ [\%]& (0.52, 2.7, 26)  & (3.0$\times 10^{-3}$, 0.016, 0.21)  \\  \hline        
Br$(H^\pm \to W^\pm \gamma)$ [\%]&(1.1, 1.5, 1.2) & (6.5$\times 10^{-3}$, 8.6$\times 10^{-3}$, 9.3$\times 10^{-3}$)   \\  \hline     
$\sigma_{S,Z}^{\text{top}} $ [fb] &(120, 210, 8.6)& (0.71, 1.3, 0.070) \\  \hline
$\sigma_{S,\gamma}^{\text{top}} $ [fb] &(260, 120, 0.40)& (1.5, 0.68, 3.1$\times 10^{-3}$) \\  \hline
$\sigma_{S,Z}^{\text{EW}} $ [fb]  &(1.8, 5.8, 36) & (0.010, 0.034, 0.29) \\  \hline
$\sigma_{S,\gamma}^{\text{EW}} $ [fb]  &(3.8, 3.2, 1.7) & (0.022, 0.018, 0.013) \\  \hline\hline
\end{tabular}}
\caption{Same as Table~\ref{res1} but for $\tan\beta=4$. }
\label{res2}
\end{center}
\end{table}

Finally, we discuss the signature of the $H^\pm \to W^\pm Z$ and $H^\pm \to W^\pm \gamma$ decays at the LHC. 
If the $H^\pm$ mass is below the top quark mass, the top decay $t \to H^\pm b$ is the dominant production mode of $H^\pm$ while above it $H^\pm$-strahlung becomes dominant. In reality, the latter is never significant as a means of
 enabling $H^\pm \to W^\pm Z$ and $H^\pm \to W^\pm \gamma$ detection, so we only concentrate on the former.
We then expect the signature $pp \to b\bar{b}H^\pm W^\mp \to b\bar{b}W^\pm W^\mp V$. 
The signal cross section of this process $\sigma_S^{\text{top}}$ is estimated by 
\begin{align}
\sigma_{S,V}^{\text{top}} 
& = 2\times \sigma_{t\bar{t}}\times [1-\text{BR}(t\to H^\pm b)]\times\text{BR}(t\to H^\pm b)\times \text{BR}(H^\pm \to W^\pm V), \label{xs_top}
\end{align}
where $\sigma_{t\bar{t}}$ is the top quark pair production cross section at the LHC. 
In Ref.~\cite{ttbar}, $\sigma_{t\bar{t}}=923.0$ pb has been obtained with $m_t=171$ GeV and $\sqrt{s}=14$ TeV 
at the next-to-next-to leading order 
using {\tt CTEQ6.6} parton distribution function~\cite{CTEQ}. 
As alternative production modes of $H^\pm$ states, especially helpful when the charged Higgs mass is larger than the
top quark mass, one should also count the EW productions, e.g., 
$pp \to H^\pm A$, $pp \to H^\pm H$ and $pp \to H^+ H^- $ whose cross sections are determined by the masses of extra Higgs bosons. 
The cross sections for $H^\pm A$ and $H^\pm H$ productions are the same as long as we take $m_A^{}=m_H^{}$ and $\sin(\beta-\alpha)=1$. 
By using these production modes, we can consider $pp \to H^\pm A/H^\pm H \to W^\pm V+X^0$ and  $pp \to H^+ H^- \to W^\pm V+X^\mp$,
where
$X^0$ and $X^\pm$ are respectively the decay product of $A/H$ and $H^\pm$. 
The signal cross section via the EW production modes are estimated by 
\begin{align}
\sigma_{S,V}^{\text{EW}} = (\sigma_{H^\pm A} + \sigma_{H^\pm H} +2\sigma_{H^+ H^-})\times \text{BR}(H^\pm \to W^\pm V), \label{xs_ew}
\end{align}
where $\sigma_{H^\pm A}$, $\sigma_{H^\pm H}$ and $\sigma_{H^+ H^-}$ are respectively 
the cross sections of $pp\to H^\pm A$, $pp\to H^\pm H$ and $pp\to H^+ H^-$. 
In the cases of $m_{H^\pm}=130$, 150 and 170 GeV, 
we obtain $\sigma_{H^\pm A}~(\sigma_{H^+ H^-})=84$~(89), 54~(53) and 36~(34) fb , 
respectively, at $\sqrt{s}=14$ TeV using {\tt CTEQ6L}. 
For $\sigma_{H^\pm A}~(=\sigma_{H^\pm H})$, the above numbers are obtained by summing the $H^+A$ and $H^-A$ processes.

In Tabs.~\ref{res1} and \ref{res2}, we show the branching fractions of the $t\to H^\pm b$, $H^\pm \to W^\pm Z$ and $H^\pm \to W^\pm \gamma$ modes and 
the overall 
signal cross sections of both the top decay and EW processes estimated by using  Eqs.~(\ref{xs_top}) and (\ref{xs_ew}), respectively. 
The results with the Type-I (X) Yukawa interaction are given in Tab.~\ref{res1} (\ref{res2}). 
For the top decay process, the production cross section gets smaller when $m_{H^\pm}$ approaches $m_t$ because of the phase space suppression. 
Conversely, the branching fraction for $H^\pm \to W^\pm Z$ becomes larger as we already seen in Figs.~\ref{BR1} and ~\ref{BRX}. 
As a result, $\sigma_{S,Z}^{\text{top}}$ attains a maximal value  around $m_{H^\pm}\simeq$ 150 GeV, while 
$\sigma_{S,\gamma}^{\text{top}}$ is simply reduced as $m_{H^\pm}^{}$ becomes larger since BR($H^\pm \to W^\pm \gamma$) does not encounter any threshold (as $m_{H^\pm}>m_{W^\pm}$).
For the EW processes, the reduction of the production cross section ($\sigma_{H^\pm A}$, $\sigma_{H^\pm H}$ and $\sigma_{H^+ H^-}$) 
is milder than that of the top decay process ($\sigma_{t\bar{t}}\times$Br($t\to H^\pm b$)). 
Therefore, the signal cross section of the EW processes become larger than the top decay process at $m_{H^\pm}=170$ GeV.  Finally,
we note that the signal cross sections in the Type-X case is more than one order of magnitude smaller than those in the Type-I case.

\section{Conclusion}\label{sec:conclusion}

We have computed the strength of the $H^\pm W^\mp Z$ and $H^\pm W^\mp \gamma$ vertices at the one-loop level in the 3HDM under a $Z_2\times\tilde{Z}_2$ symmetry, which defines a Higgs sector with two active doublets and one inert one. 
We have discussed all the four types of the Yukawa interactions which are defined by the $\tilde{Z}_2$ charge assignment to the SM fermions. 
We have taken into account vacuum stability and perturbative unitarity as theoretical constraints, and have considered 
the bounds from the EW $S$, $T$ and $U$ parameters, flavour experiments and direct searches for $H^\pm $ states at LEP-II and LHC Run-I. 
We have seen that the mass of the $H^\pm$ can be smaller than the top quark mass in models with the
Type-I and Type-X Yukawa interactions, but not in Type-II and Type-Y. Further, we have shown that,
among all the form factors, only $F_Z$ can be enhanced with respect to the 2HDM 
by taking large mass splittings between $\eta_A^{}$ and $\eta^\pm$, 
because of the non-decoupling effect of the inert scalar boson loop contributions.

In particular, we have found that in the 3HDM  
the squared form factor $|F_Z|^2$ can be one order of magnitude larger than that predicted in the 2HDM under the aforementioned theoretical and experimental constraints. 
In addition, the branching fraction of the $H^\pm \to W^\pm Z$ mode
can be about 4~(0.2)\%, 40~(2)\% and 0.4~(0.3)\% in the cases of $m_{H^\pm}=150$, 170 and 200 GeV, respectively with the Type-I (Type-X) Yukawa interactions.  
In contrast, the branching fraction of the $H^\pm \to W^\pm \gamma$ mode is at the few percent level as long as $m_{H^\pm}$ is smaller than the top quark mass in the Type-I and Type-X cases, thus benefiting from very little
enhancement with respect to the 2HDM.  Such increased rates in the 3HDM stem from loop contributions due to inert
Higgs states that are absent in the 2HDM.

Finally, we have discussed signal processes embedding 
$H^\pm\to W^\pm Z$ and $H^\pm\to W^\pm \gamma$ decays 
at the LHC.  In the light $H^\pm$ scenario, i.e.,  $m_{H^\pm}<m_t$, with the Type-I and Type-X Yukawa interactions,
 the top quark decay process $t\to H^\pm b$ is  the dominant production mode for $H^\pm$ except for the extreme 
case of $m_{H^\pm}\lesssim m_t$.  In the heavy  $H^\pm$ scenario, i.e.,  $m_{H^\pm}>m_t$, this channel is
no longer viable and we have resorted to the $bg\to tH^\pm$ mode. (Herein, we have emulated top production plus decay
and $H^\pm$-strahlung via $gg\to tbH^\pm$.) In fact, 
there are also EW production modes, such as $pp\to H^+ H^-$, $pp\to H^\pm A$ and $pp\to H^\pm H$. 
By combining the production and decay of $H^\pm$'s, we have considered 
the signal processes $pp\to b\bar{b}W^\pm H^\mp \to b\bar{b} W^+ W^- V$, 
$pp\to H^+H^- \to W^\pm V X^\mp$ and $pp\to H^\pm A/H^\pm H  \to W^\pm V X^0$. 
We have thus computed the ensuing cross sections in all cases and shown that the LHC Run-II has the potential
to access $H^\pm\to W^\pm Z$ and/or $H^\pm\to W^\pm \gamma$ decays, certainly for light $H^\pm$'s (at standard
luminosity) and possibly 
for heavy $H^\pm$'s (at very high luminosity). To establish one or the other such signals at the CERN machine may represent circumstantial
evidence of a 3HDM sector, as opposed to a 2HDM.

\section*{Acknowledgments}

S. M. is supported in part through the NExT Institute. D. R. is financed in part by CONACYT-M\'exico. 
 K.~Y. is fully supported by a JSPS postdoctoral fellowships for research abroad.

\clearpage

\begin{appendix}

\section{1PI contributions}\label{sec:1pi}

\begin{figure}[h]
\begin{center}
\includegraphics[width=130mm]{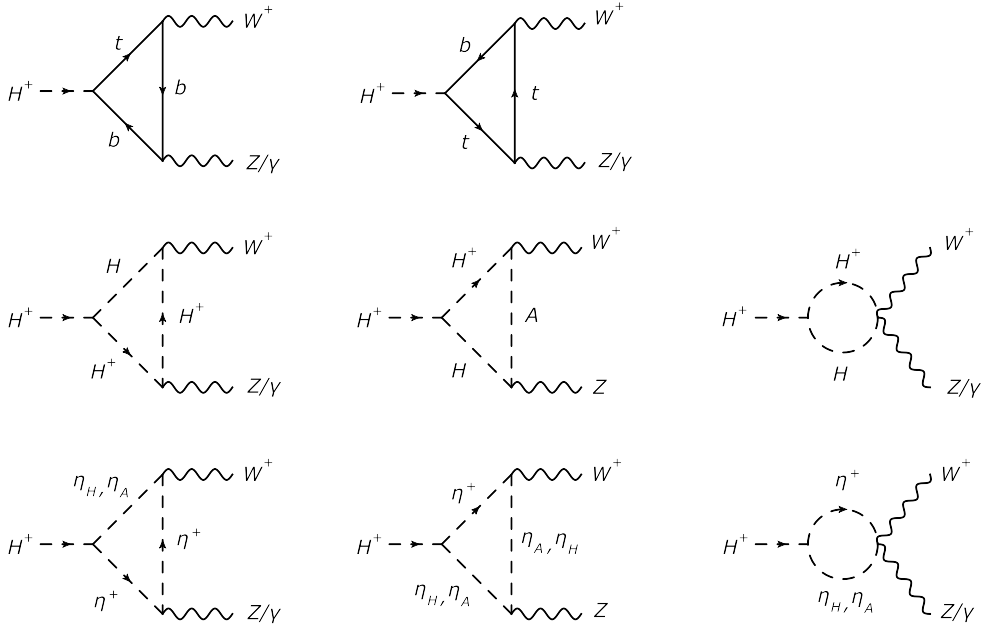}
\caption{The 1PI diagrams for the $H^\pm W^\mp Z$ and $H^\pm W^\mp \gamma$ vertices.  
The diagrams which vanish in the limit  $\sin(\beta-\alpha)=1$ are not displayed.  }
\label{fig_1pi1}
\end{center}
\end{figure}

Here, we give the analytic expressions for the 1PI diagram contributions to the form factors of the $H^\pm W^\mp V$ ($V=Z,~\gamma$) vertices
and those for the $W^\pm$-$H^\pm$ mixing $\Gamma_{WH}^{\text{1PI}}$.   
The fermion loop contribution to the $H^\pm W^\mp V$ vertices has been calculated in Ref.~\cite{f}
whereas the boson contribution in the 2HDM has been evaluated in Refs.~\cite{b1,b2}. 
In addition to these contributions, there are inert scalar boson loop contributions as shown in Fig.~\ref{fig_1pi1}. 

In the following, we separately show the fermion and boson loop contributions to the form factors denoted by 
$X_{V,F}^{\text{1PI}}$ and $X_{V,B}^{\text{1PI}}$ ($X=F,G$ and $H$), respectively. 
Regarding the boson loop contribution, we only show the contributions from pure scalar loop diagrams, where 
scalar bosons are running in the triangle and circle type diagrams (see Fig.~\ref{fig_1pi1}). 
There are additional gauge-scalar mixed type diagrams, where one gauge and two active scalar bosons or 
two gauge and one active scalar bosons run in the triangle part.  
Because these contributions are proportional to $\cos(\beta-\alpha)$, 
they vanish or become negligible by taking the SM-like limit $\sin(\beta-\alpha)\to 1$ or taking the SM-like regime $\sin(\beta-\alpha)\simeq 1$, respectively.  
We thus neglect them here\footnote{The 
contributions from the gauge-scalar mixed type diagrams are given in Ref.~\cite{b1}. }. 

\begin{figure}[h]
\begin{center}
\includegraphics[width=130mm]{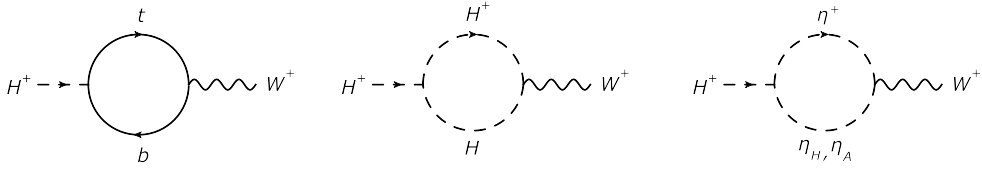}
\caption{Diagrams giving the $H^\pm$-$W^\mp$ mixing.  
The diagrams which vanish in the limit  $\sin(\beta-\alpha)=1$ are not displayed.  }
\label{fig_1pi2}
\end{center}
\end{figure}

In order to express loop functions, we use the Passarino-Veltman functions~\cite{PV}. 
Here, we give the integral formulae of some of the functions which we use in the following discussion:
\begin{subequations}
\begin{align}
B_0(p^2;m_1,m_2)&=\Delta -\int_0^1 dx \ln \Delta_B, \\
B_1(p^2;m_1,m_2) &=-\frac{\Delta}{2}+\int_0^1 dx (1-x)\ln \Delta_B, \\
C_0(p_1^2,p_2^2,q^2;m_1,m_2,m_3)&=-\int_0^1 dx \int_0^1dy\frac{y}{\Delta_C},\\ 
C_{11}(p_1^2,p_2^2,q^2;m_1,m_2,m_3)&=-\int_0^1 dx \int_0^1dy\frac{y(xy-1)}{\Delta_C},\\
C_{12}(p_1^2,p_2^2,q^2;m_1,m_2,m_3)&=-\int_0^1 dx \int_0^1dy\frac{y(y-1)}{\Delta_C}, \\
C_{21}(p_1^2,p_2^2,q^2;m_1,m_2,m_3)&=-\int_0^1 dx \int_0^1dy\frac{y(1-xy)^2}{\Delta_C},\\
C_{22}(p_1^2,p_2^2,q^2;m_1,m_2,m_3)&=-\int_0^1 dx \int_0^1dy\frac{y(1-y)^2}{\Delta_C},\\
C_{23}(p_1^2,p_2^2,q^2;m_1,m_2,m_3)&=-\int_0^1 dx \int_0^1dy\frac{y(1-xy)(1-y)}{\Delta_C}, \\ 
C_{24}(p_1^2,p_2^2,q^2;m_1,m_2,m_3)&=\frac{\Delta}{4} -\frac{1}{2}\int_0^1 dx \int_0^1dy \,y\ln \Delta_C, 
\end{align}
\label{pv_int}
\end{subequations}
where
\begin{align}
\Delta_B  &= -x(1-x)p^2 + xm_1^2 + (1-x)m_2^2, \\
\Delta_C &= y^2(p_1x+p_2)^2+y[x(p_2^2-q^2+m_1^2-m_2^2)+m_2^2-m_3^2-p_2^2]+m_3^2, 
\end{align}
In Eq.~(\ref{pv_int}), $\Delta$ is given by
\begin{align}
\Delta \equiv \frac{1}{\epsilon}-\gamma_E+\ln 4\pi+\ln\mu^2, \label{div}
\end{align} 
where $\epsilon$ appears in the $D(=4-2\epsilon)$ dimensional integral, $\mu$ is an arbitrary dimensionful parameter and  $\gamma_E^{}$ is the Euler constant. 
In the four dimension limit $\epsilon\to 0$, $\Delta$ is divergent. 
We note that this divergent part $\Delta$ appears in the following expressions, but it is exactly cancelled in the renormalized variables such as 
$X_Z$ and $X_\gamma$ ($X=F,~G$ and $H$). 
We use the shorthand notations like $B_{i}(p^2;A,B)=B_{i}(p^2;m_A,m_B)$ and $C_{i,~ij}(A,B,C)=C_{i,~ij}(p_1^2,p_2^2,q^2;m_A,m_B,m_C)$.

The fermion loop contribution to $X_Z^{\text{1PI}}$ is given by 
\begin{align}
&      F_{Z,F}^{\text{1PI}}
= \frac{2N_c}{16\pi^2 v^2c_W^{}} \Big\{\notag\\
&+m_t^2\xi_t (v_b+a_b)\Big[4C_{24}(t,b,b)-B_0(q^2;m_t,m_b)-B_0(p_W^2;m_b,m_t)-(2m_b^2-p_Z^2)C_0(t,b,b) \Big]\notag\\
&-m_b^2\xi_b(v_b+a_b)\Big[4C_{24}(t,b,b)-B_0(p_Z^2;m_b,m_b)-B_0(q^2;m_t,m_b)-(m_t^2+m_b^2-p_W^2)C_0(t,b,b)\Big]\notag\\
&-m_b^2\xi_b(v_b-a_b)\Big[B_0(p_Z^2;m_b,m_b)+B_0(p_W^2;m_t,m_b)+(m_t^2+m_b^2-q^2)C_0(t,b,b)\Big]\notag\\
&+2m_t^2m_b^2\xi_t(v_b-a_b)C_0(t,b,b)
\Big\} 
+(m_t,\xi_t,v_b,a_b) \leftrightarrow (m_b,-\xi_b,v_t,a_t), \\
&G_{Z,F}^{\text{1PI}}
 = \frac{4N_cm_W^2}{16\pi^2 v^2c_W^{}} \Big[
m_t^2\xi_t (v_b+a_b)(2C_{23}+2C_{12}+C_{11}+C_0 )\notag\\
&-m_b^2\xi_b(v_b+a_b)(2C_{23} + C_{12})
-m_b^2\xi_b(v_b-a_b)(C_{12}-C_{11})\Big](t,b,b) \notag\\
&+(m_t,\xi_t,v_b,a_b) \leftrightarrow (m_b,-\xi_b,v_t,a_t), \\
&H_{Z,F}^{\text{1PI}}
 = \frac{4N_cm_W^2}{16\pi^2 v^2c_W^{}} \Big[
m_t^2\xi_t (v_b+a_b)(C_{0}+C_{11})
-m_b^2\xi_b(v_b+a_b)C_{12}
+m_b^2\xi_b(v_b-a_b)(C_{12}-C_{11})\Big](t,b,b) \notag\\
&+(m_t,\xi_t,v_b,a_b) \leftrightarrow (m_b,+\xi_b,v_t,a_t), 
\end{align}
where 
\begin{align}
v_f = \frac{1}{2}I_f-s_W^2 Q_f,\quad a_f = \frac{1}{2}I_f. 
\end{align}
That to $X_\gamma^{\text{1PI}}$ is given by 
\begin{align}
&      F_{\gamma,F}^{\text{1PI}}
= \frac{2N_cQ_b}{16\pi^2 v^2c_W^{}} \Big\{\notag\\
&+m_t^2\xi_t \Big[4C_{24}(t,b,b)-B_0(q^2;m_t,m_b)-B_0(p_W^2;m_b,m_t)-(2m_b^2-p_\gamma^2)C_0(t,b,b) \Big]\notag\\
&-m_b^2\xi_b\Big[4C_{24}(t,b,b)-B_0(p_Z^2;m_b,m_b)-B_0(q^2;m_t,m_b)-(m_t^2+m_b^2-p_W^2)C_0(t,b,b)\Big]\notag\\
&-m_b^2\xi_b\Big[B_0(p_Z^2;m_b,m_b)+B_0(p_W^2;m_t,m_b)+(m_t^2+m_b^2-q^2)C_0(t,b,b)\Big]\notag\\
&+2m_t^2m_b^2\xi_tC_0(t,b,b)
\Big\} 
+(m_t,\xi_t,Q_b) \leftrightarrow (m_b,-\xi_b,Q_t), \\
&G_{\gamma,F}^{\text{1PI}}
 = \frac{4N_cQ_bm_W^2}{16\pi^2 v^2c_W^{}} \Big[
m_t^2\xi_t (2C_{23}+2C_{12}+C_{11}+C_0 )\notag\\
&-m_b^2\xi_b(2C_{23} + C_{12})
-m_b^2\xi_b(C_{12}-C_{11})\Big](t,b,b)
+(m_t,\xi_t,Q_b) \leftrightarrow (m_b,-\xi_b,Q_t), \\
&H_{Z,F}^{\text{1PI}}
 = \frac{4N_cQ_bm_W^2}{16\pi^2 v^2c_W^{}} \Big[
m_t^2\xi_t (C_{0}+C_{11})
-m_b^2\xi_bC_{12}
+m_b^2\xi_b(C_{12}-C_{11})\Big](t,b,b) \notag\\
&+(m_t,\xi_t,Q_b) \leftrightarrow (m_b,+\xi_b,Q_t). 
\end{align}
The boson loop contribution is given by 
\begin{align}
&F_{Z,B}^{\text{1PI}}=\frac{1}{16 \pi^2 v c_W^{}}\Big\{\notag\\
&+\lambda_{H^+H^-H}
  \sin(\beta-\alpha)\Big[(2-4s_W^2)C_{24}(H,H^\pm,H^\pm)-2C_{24}(H^\pm, A,H)+s_W^2B_0 (q^2;H^\pm,H) \Big]  \notag\\
&- \lambda_{H^+\eta^- \eta_H^{}}\Big[(2-4s_W^2)C_{24}(\eta_H^{},\eta^\pm,\eta^\pm)-2C_{24}(\eta^\pm, \eta_A^{},\eta_H^{})+s_W^2B_0 (q^2; \eta^\pm,\eta_H^{}) \Big]  \notag\\
&- \lambda_{H^+\eta^- \eta_A^{}} \Big[(2-4s_W^2)C_{24}(\eta_A^{},\eta^\pm,\eta^\pm)-2C_{24}(\eta^\pm, \eta_H^{},\eta_A^{})+s_W^2B_0 (q^2;\eta^\pm,\eta_A^{}) \Big]   \Big\}, \\
&G_{Z,B}^{\text{1PI}}=\frac{m_W^2}{16 \pi^2 v c_W^{}}\Big\{\notag\\
&+\lambda_{H^+H^-H}
  \sin(\beta-\alpha)\Big[(2-4s_W^2)(C_{12}+C_{23})(H,H^\pm,H^\pm)-2(C_{12}+C_{23})(H^\pm, A,H) \Big]  \notag\\
&- \lambda_{H^+\eta^- \eta_H^{}}\Big[(2-4s_W^2)(C_{12}+C_{23})(\eta_H^{},\eta^\pm,\eta^\pm)-2(C_{12}+C_{23})(\eta^\pm, \eta_A^{},\eta_H^{}) \Big]  \notag\\
&- \lambda_{H^+\eta^- \eta_A^{}} \Big[(2-4s_W^2)(C_{12}+C_{23})(\eta_A^{},\eta^\pm,\eta^\pm)-2(C_{12}+C_{23})(\eta^\pm, \eta_H^{},\eta_A^{}) \Big]   \Big\},  \\
&F_{\gamma,B}^{\text{1PI}}=\frac{s_W^{}}{16 \pi^2 v}\Big\{
\lambda_{H^+H^-H}
  \sin(\beta-\alpha)[4C_{24}(H,H^\pm,H^\pm)-B_0 (q^2;H^\pm,H) ]  \notag\\
&- \lambda_{H^+\eta^- \eta_H^{}}[4C_{24}(\eta_H^{},\eta^\pm,\eta^\pm)-B_0 (q^2; \eta^\pm,\eta_H^{}) ]  
- \lambda_{H^+\eta^- \eta_A^{}} [4C_{24}(\eta_A^{},\eta^\pm,\eta^\pm)-B_0 (q^2;\eta^\pm,\eta_A^{}) ]   \Big\}, \\
&G_{\gamma,B}^{\text{1PI}}=\frac{4m_W^2s_W^{}}{16 \pi^2 v }\Big[\lambda_{H^+H^-H}
  \sin(\beta-\alpha)(C_{12}+C_{23})(H,H^\pm,H^\pm)\notag\\
& - \lambda_{H^+\eta^- \eta_H^{}}(C_{12}+C_{23})(\eta_H^{},\eta^\pm,\eta^\pm)
- \lambda_{H^+\eta^- \eta_A^{}}(C_{12}+C_{23})(\eta_A^{},\eta^\pm,\eta^\pm)    \Big], 
\end{align}
and 
\begin{align}
H_{Z,B}^{\text{1PI}}=H_{\gamma,B}^{\text{1PI}}=0,  
\end{align}
where 
\begin{align}
\lambda_{H^+H^-H}&=
\frac{1}{v}\Big[(m_H^2-M^2)(\cot\beta-\tan\beta)\sin(\beta-\alpha)-(2m_{H^\pm}^2+m_H^2-2M^2)\cos(\beta-\alpha)\Big], \label{HpHmH}  \\
\lambda_{H^\pm \eta^\mp \eta_H^{}} &= \frac{v}{4}(\rho_2+\rho_3-\sigma_2-\sigma_3)\sin 2\beta, \label{epemH} \\
\lambda_{H^\pm \eta^\mp \eta_A^{}} &= \pm\frac{v}{4}(\rho_2-\rho_3-\sigma_2+\sigma_3)\sin 2\beta. \label{epemA}   
\end{align}
We note that the above expressions are obtained by extracting the coefficient of the scalar trilinear vertex, i.e., 
$\mathcal{L} = +\lambda_{\phi_1\phi_2\phi_3}^{}\phi_1\phi_2\phi_3\,  + \cdots $. 

The fermion and boson loop contributions to the $W^\pm$-$H^\pm$ mixing, i.e., $\Gamma_{WH}^{\text{1PI}}(p^2)_F$ and $\Gamma_{WH}^{\text{1PI}}(p^2)_B$, respectively,  
are given by:  
\begin{align}
\Gamma_{WH}^{\text{1PI}}(p^2)_F &= \frac{i}{16\pi^2}\frac{4m_W^{}}{v^2}N_c[m_t^2 \xi_t (B_0+B_1)-m_b^2\xi_b B_1](p^2;m_t,m_b), \\ 
\Gamma_{WH}^{\text{1PI}}(p^2)_B &= \frac{i}{16\pi^2}\frac{m_W^{}}{v}\Big[\lambda_{H^+H^-H}\sin(\beta-\alpha) (2B_1+B_0)(p^2;m_{H^\pm}^{},m_H^{})\notag\\
&+\lambda_{\eta^+\eta^-\eta_H}(2B_1+B_0)(p^2;m_{\eta^\pm}^{},m_{\eta_H}^{})
+i\lambda_{\eta^+\eta^-\eta_A}(2B_1+B_0)(p^2;m_{\eta^\pm}^{},m_{\eta_A}^{})
\Big]. 
\end{align}
The counter term contribution is then obtained from the above $W^\pm$-$H^\pm$ mixing via Eq.~(\ref{delFF}):  
\begin{align}
&\delta F_{Z,F} = \frac{4s_W^2N_c}{16\pi^2v^2c_W^{}}[m_t^2 \xi_t (B_0+B_1)-m_b^2\xi_b B_1](q^2;t,b), \\
&\delta F_{Z,B}= \frac{s_W^2}{16\pi^2v c_W^{}}\Big[\lambda_{H^+H^-H}\sin(\beta-\alpha) (2B_1+B_0)(q^2;H^\pm,H)\notag\\
&-\lambda_{H^+\eta^-\eta_H}(2B_1+B_0)(q^2;\eta^\pm,\eta_H^{})
-\lambda_{H^+\eta^-\eta_A}(2B_1+B_0)(q^2;\eta^\pm,\eta_A^{})
\Big], \\
&\delta F_{\gamma,F/B} = -\frac{c_W^{}}{s_W^{}}\delta F_{Z,F/B}. 
\end{align}

Using the above analytic expressions, we can directly check the relation from the Ward identity in Eq.~(\ref{ward}), i.e., 
$(F_\gamma^{\text{1PI}}+\delta F_\gamma)=G_\gamma^{\text{1PI}}(1-m_W^2/m_{H^\pm}^2)/2$.

\end{appendix}                  %


\begin{thebibliography}{1}

\bibitem{Higgs_coupling1} 
  G.~Aad {\it et al.}  [ATLAS Collaboration],
  Phys.\ Lett.\ B {\bf 726}, 88 (2013)
  [Erratum-ibid.\ B {\bf 734}, 406 (2014)]. 

\bibitem{Higgs_coupling2} 
  G.~Aad {\it et al.}  [ATLAS Collaboration],
  Phys.\ Rev.\ D {\bf 91}, 012006 (2015). 

\bibitem{Higgs_coupling3}
  S.~Chatrchyan {\it et al.}  [CMS Collaboration],
  JHEP {\bf 1401}, 096 (2014). 

\bibitem{Higgs_coupling4} 
  V.~Khachatryan {\it et al.}  [CMS Collaboration],
  arXiv:1412.8662 [hep-ex].



\bibitem{HHG}
 
  J.~F.~Gunion, H.~E.~Haber, G.~L.~Kane and S.~Dawson,
  Front.\ Phys.\  {\bf 80}, 1 (2000).


\bibitem{Zee}
See, e.g.,  A.~Zee,
 Phys.\ Lett.\  B {\bf 93}, 389 (1980) 
 [Erratum-ibid.\  B {\bf 95}, 461  (1980)].


\bibitem{IDM} 
  R.~Barbieri, L.~J.~Hall and V.~S.~Rychkov,
  Phys.\ Rev.\ D {\bf 74}, 015007 (2006).

\bibitem{CPV} 
  T.~D.~Lee,
  Phys.\ Rev.\ D {\bf 8}, 1226 (1973); 

  S.~Weinberg,
  Phys.\ Rev.\ Lett.\  {\bf 37}, 657 (1976).

\bibitem{Grifols} 
  J.~A.~Grifols and A.~Mendez,
  Phys.\ Rev.\ D {\bf 22}, 1725 (1980).

\bibitem{f} 
  M.~Capdequi Peyranere, H.~E.~Haber and P.~Irulegui,
  Phys.\ Rev.\ D {\bf 44}, 191 (1991).

\bibitem{b1} 
  S.~Kanemura,
  Phys.\ Rev.\ D {\bf 61}, 095001 (2000).

\bibitem{b2} 
  S.~Kanemura,
  Eur.\ Phys.\ J.\ C {\bf 17}, 473 (2000).

\bibitem{HWZ_LHC}
  E.~Asakawa and S.~Kanemura,
  Phys.\ Lett.\ B {\bf 626}, 111 (2005);

  E.~Asakawa, S.~Kanemura and J.~Kanzaki,
  Phys.\ Rev.\ D {\bf 75}, 075022 (2007);

  S.~Godfrey and K.~Moats,
  Phys.\ Rev.\ D {\bf 81}, 075026 (2010). 

\bibitem{Yanase} 
  S.~Kanemura, K.~Yagyu and K.~Yanase,
  Phys.\ Rev.\ D {\bf 83}, 075018 (2011).


\bibitem{idm2} 
  B.~Grzadkowski, O.~M.~Ogreid, P.~Osland, A.~Pukhov and M.~Purmohammadi,
  JHEP {\bf 1106}, 003 (2011)
  [arXiv:1012.4680 [hep-ph]].

\bibitem{3hdm_vs} 
  B.~Grzadkowski, O.~M.~Ogreid and P.~Osland,
  Phys.\ Rev.\ D {\bf 80}, 055013 (2009). 

\bibitem{2plus1_1} 
  V.~Keus, S.~F.~King and S.~Moretti,
  Phys.\ Rev.\ D {\bf 90}, no. 7, 075015 (2014). 


\bibitem{2plus1_2} 
  V.~Keus, S.~F.~King, S.~Moretti and D.~Sokolowska,
  JHEP {\bf 1411}, 016 (2014).

\bibitem{GW}
  S.~L.~Glashow, S.~Weinberg,
  Phys.\ Rev.\ D {\bf 15 }, 1958 (1977).


\bibitem{4types}

  V.~D.~Barger, J.~L.~Hewett and R.~J.~N.~Phillips,
  Phys.\ Rev.\ D {\bf 41}, 3421 (1990);

  Y.~Grossman,
  Nucl.\ Phys.\ B {\bf 426}, 355 (1994).

\bibitem{4types2}

  A.~G.~Akeroyd,
  Phys.\ Lett.\ B {\bf 377}, 95 (1996).


\bibitem{typeX} 

  M.~Aoki, S.~Kanemura, K.~Tsumura and K.~Yagyu,
  Phys.\ Rev.\ D {\bf 80}, 015017 (2009).

\bibitem{Gunion-Haber}
  J.~F.~Gunion and H.~E.~Haber,
  Phys.\ Rev.\ D {\bf 67}, 075019 (2003).

\bibitem{Dev} 
  P.~S.~B.~Dev and A.~Pilaftsis,
  JHEP {\bf 1412}, 024 (2014).

\bibitem{abe} 
  T.~Abe and R.~Kitano,
  Phys.\ Rev.\ D {\bf 88}, no. 1, 015019 (2013). 


\bibitem{Pomarol} 
  A.~Pomarol and R.~Vega,
  Nucl.\ Phys.\ B {\bf 413}, 3 (1994). 


\bibitem{KOSY}
  S.~Kanemura, Y.~Okada, E.~Senaha and C.~-P.~Yuan,
  Phys.\ Rev.\ D {\bf 70}, 115002 (2004).

\bibitem{finger2} 
  S.~Kanemura, M.~Kikuchi and K.~Yagyu,
  arXiv:1502.07716 [hep-ph].



\bibitem{Moretti-Yagyu} 
  S.~Moretti and K.~Yagyu,
  Phys.\ Rev.\ D {\bf 91}, 055022 (2015). 


\bibitem{stu}
  M.~E.~Peskin and T.~Takeuchi,
  Phys.\ Rev.\ Lett.\  {\bf 65}, 964 (1990) and 
  Phys.\ Rev.\  D {\bf 46}, 381 (1992). 


  \bibitem{Baak:2012kk}
  M.~Baak, M.~Goebel, J.~Haller, A.~Hoecker, D.~Kennedy, R.~Kogler, K.~Moenig and M.~Schott {\it et al.},
  Eur.\ Phys.\ J.\ C {\bf 72}, 2205 (2012).

\bibitem{bsg_322} 
  O.~Eberhardt, U.~Nierste and M.~Wiebusch,
  JHEP {\bf 1307}, 118 (2013). 

\bibitem{Misiak}
  T.~Hermann, M.~Misiak and M.~Steinhauser,
  JHEP {\bf 1211}, 036 (2012). 


\bibitem{Stal}
  F.~Mahmoudi and O.~Stal,
  Phys.\ Rev.\ D {\bf 81}, 035016 (2010). 

\bibitem{LEPII} 
  G.~Abbiendi {\it et al.}  [ALEPH and DELPHI and L3 and OPAL and LEP Collaborations],
  Eur.\ Phys.\ J.\ C {\bf 73}, 2463 (2013).

\bibitem{ATLAS_H+} 

  G.~Aad {\it et al.}  [ATLAS Collaboration],
  arXiv:1412.6663 [hep-ex].


\bibitem{Guchait:2001pi} 
  M.~Guchait and S.~Moretti,
  JHEP {\bf 0201}, 001 (2002).

\bibitem{cb}
  A.~G.~Akeroyd, S.~Moretti and J.~Hernandez-Sanchez,
  Phys.\ Rev.\ D {\bf 85}, 115002 (2012);
 
 arXiv:1409.7596 [hep-ph].
 
\bibitem{PDG2014} 
K.A.~Olive et al.~(Particle Data Group), Chin.~Phys.~C, 38, 090001 (2014). 

\bibitem{Koide} 
 H.~Fusaoka and Y.~Koide,
 Phys.\ Rev.\ D {\bf 57}, 3986 (1998). 

 \bibitem{ttbar} 

  U.~Langenfeld, S.~Moch and P.~Uwer,
  Phys.\ Rev.\ D {\bf 80}, 054009 (2009).


\bibitem{CTEQ} 
  P.~M.~Nadolsky, H.~L.~Lai, Q.~H.~Cao, J.~Huston, J.~Pumplin, D.~Stump, W.~K.~Tung and C.-P.~Yuan,
  Phys.\ Rev.\ D {\bf 78}, 013004 (2008).



\bibitem{PV}
  G.~Passarino and M.~J.~G.~Veltman,
  Nucl.\ Phys.\  B {\bf 160}, 151 (1979).






\end{thebibliography}
\end{document}